\documentclass[letterpaper,twocolumn,10pt]{article}
\usepackage{usenix,epsfig,endnotes}

\usepackage{tikz}
\usepackage{multirow}
\usepackage{booktabs}
\usepackage{float}
\usepackage[misc]{ifsym}

\usepackage{url}

\usepackage{tcolorbox}
\tcbuselibrary{skins, breakable, theorems}
\usepackage[font=small]{caption}

\usepackage{amssymb}
\usepackage{subfigure}
\usepackage{threeparttable}

\usepackage{titlesec}
\titlespacing\section{0pt}{8pt plus 5pt minus 3pt}{2pt plus 2pt minus 3pt}
\titlespacing\subsection{0pt}{8pt plus 5pt minus 3pt}{2pt plus 2pt minus 3pt}

\newcommand{\ignore}[1]{}

\newcommand{\paragraphbe}[1]{\smallskip\noindent{\bf {#1}.}~}

\newtcolorbox{mybox}{colframe =  blue!40!white, colback = blue!10!white}

\newcommand\blfootnote[1]{%
\begingroup
\renewcommand\thefootnote{}\footnote{#1}%
\addtocounter{footnote}{-1}%
\endgroup
}
\newcommand*\emptycirc[1][1ex]{\tikz\draw (0,0) circle (#1);} 
\newcommand*\halfcirc[1][1ex]{%
	\begin{tikzpicture}
	\draw[fill] (0,0)-- (90:#1) arc (90:270:#1) -- cycle ;
	\draw (0,0) circle (#1);
	\end{tikzpicture}}
\newcommand*\fullcirc[1][1ex]{\tikz\fill (0,0) circle (#1);}

\def\BibTeX{{\rm B\kern-.05em{\sc i\kern-.025em b}\kern-.08em
    T\kern-.1667em\lower.7ex\hbox{E}\kern-.125emX}}

\begin{document}





\title{\Large \bf Seeing is Living? Rethinking the Security of Facial Liveness Verification \\in the Deepfake Era }

\pagestyle{empty}

\author{
{\rm Changjiang Li}$^{\dagger \ddagger}$ \quad {\rm Li Wang}$^\S$ \quad {\rm Shouling Ji}$^\ddagger $ \textsuperscript{\Letter} \quad {\rm Xuhong Zhang}$^\ddagger$ \textsuperscript{\Letter}\\ {\rm Zhaohan Xi}$^\dagger$ \quad {\rm Shanqing Guo}$^\S$ \quad {\rm Ting Wang}$^\dagger$\\
$^\dagger$Pennsylvania State University \quad  $^\ddagger$Zhejiang University \quad  $^\S$Shandong University
} 

\maketitle

\begin{abstract}

	Facial Liveness Verification (FLV) is widely used for identity authentication in many security-sensitive domains and offered as Platform-as-a-Service (PaaS) by leading cloud vendors. Yet, with the rapid advances in synthetic media techniques (e.g., deepfake), the security of FLV is facing unprecedented challenges, about which little is known thus far. 
	

	To bridge this gap, in this paper, we conduct the first systematic study on the security of FLV in real-world settings. Specifically, we present \texttt{LiveBugger}, a new deepfake-powered attack framework that enables customizable, automated security evaluation of FLV.  
	Leveraging \texttt{LiveBugger}, we perform a comprehensive empirical assessment of representative FLV platforms, leading to a set of interesting findings. For instance, most FLV APIs do not use anti-deepfake detection; even for those with such defenses, their effectiveness is concerning (e.g., it may detect high-quality synthesized videos but fail to detect low-quality ones).
	We then conduct an in-depth analysis of the factors impacting the attack performance of \texttt{LiveBugger}:  a) the bias (e.g., gender or race) in FLV can be exploited to select victims; b) adversarial training makes deepfake more effective to bypass FLV; c) the input quality has a varying influence on different deepfake techniques to bypass FLV. Based on these findings, we propose a customized, two-stage approach that can boost the attack success rate by up to 70\%. Further, we run proof-of-concept attacks on several representative applications of FLV (i.e., the clients of FLV APIs) to illustrate the practical implications: due to the vulnerability of the APIs, many downstream applications are vulnerable to deepfake. 
	Finally, we discuss potential countermeasures to improve the security of FLV. Our findings have been confirmed by the corresponding vendors.  

\end{abstract}

\blfootnote{Changjiang Li and Li Wang are the co-first authors. This work was partially conducted when Changjiang Li was at Zhejiang University. Shouling Ji and Xuhong Zhang are the co-corresponding authors.}

\section{Introduction}

\begin{figure}[t]
	\centerline{\includegraphics[width=0.9\columnwidth]{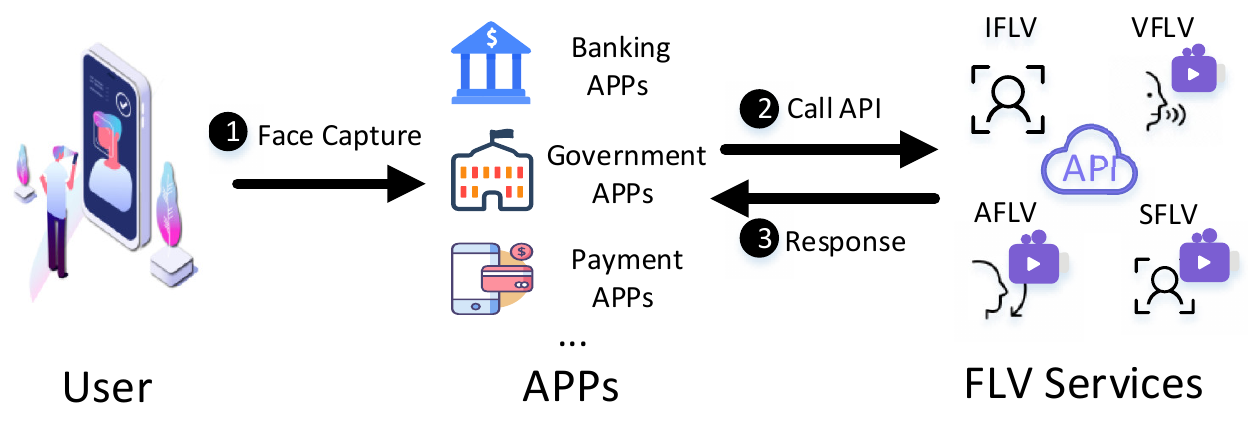}}
	\caption{\small Overview of FLV. Generally, cloud platforms provide four types of FLV including Image- (IFLV), Silence- (SFLV), Voice- (VFLV), and Action-based FLV (AFLV). }	
	\label{figure:online}
\end{figure}

As a promising alternative to legacy passwords, Facial Liveness Verification (FLV), which is able to validate an identity based on a facial image/video, has drawn increasing attention \cite{ming2018facelivenet, ivanova2020improving, smith2018continuous}. In particular, 
online-FLV 
is widely adopted in practice due to its low hardware requirement for end-users \cite{yang2018empirical}.
As shown in Figure \ref{figure:online}, it first requires the user to record a specific facial image/video, which is then sent to an FLV API for verification.
FLV has been applied in many critical scenarios, such as online payment, online banking, and government services \cite{scenarios}. For instance, recently, OCBC Bank, the second-largest bank in Singapore, has introduced FLV to eight of its ATMs as an initial trial for a larger roll-out planned throughout 2021 \cite{ocbcbank}. Besides, an increasing number of cloud platforms begin to provide FLV as Platform-as-a-Service (PaaS), which significantly reduces the cost and lowers the barrier for companies to deploy FLV in their products. These services usually  provide  APIs for downstream APPs to integrate FLV.  It is expected that the trend will continue growing at a rate of 67.6\% and lead to a \$16.6 billion market share by 2024 \cite{market}. 



In contrast to its surging popularity, the potential security risks of FLV are fairly under-explored. Especially for the APIs provided by FLV PaaS vendors, their vulnerability will be inherited by downstream APPs, threatening millions of end-users. Once the security of FLV API is compromised, the adversary may exploit it in numerous downstream APPs. The user authentication API is ranked second in ``OWASP API security top 10'' \cite{apisecurity}. Meanwhile, most existing studies  focus on the presentation attack (i.e., the replay attack), in which the adversary attempts to impersonate an identity through replaying the victim's facial image/video \cite{ramachandra2017presentation}, with strategies including printing \cite{boulkenafet2015face}, video replay \cite{li2016generalized}, and 3D mask \cite{liu2018remote}. 
In response, various defenses have also been proposed to mitigate such attacks \cite{tan2010face,maatta2011face,boulkenafet2015face,george2019biometric, wang2020cross}.
However, they study the security risk of FLV  from the algorithmic level on public datasets, without considering such risks in the deployed FLV services or systems (e.g., FLV APIs), and thus are inadequate to reflect the threat in the real-world setting.

Further, with the rapid advances in synthetic media techniques (e.g., deepfake) \cite{mirsky2021creation}, the security of FLV is facing unprecedented challenges. Although previous work (e.g., \cite{UzunCEL18}) shows that several face recognition systems are vulnerable to synthesis attacks, the vulnerability of liveness verification is largely unexplored. Further, the attack-defense landscape of FLV has changed significantly recently. Little is known about the new threats raised by state-of-the-art (SOTA)  deepfake techniques.
First, it enables more advanced and flexible attacks \cite{buo2020emerging}. For instance, it allows the adversary to easily synthesize videos with required head/lip movements based on a single image of the victim; in comparison, finding existing videos satisfying the move/voice requirements is extremely difficult in the presentation attack. In addition, with the increasing commoditization of deepfake techniques (e.g., ZAO \cite{zao}), it now requires little expertise to create fake images/videos.
 For example, recently, a group of tax scammers hacked a government-run FLV system via open-sourced deepfake techniques to fake tax invoices, which were valued at \$76.2 million \cite{taxhacker}.

Thus, it is imperative to assess the security implications of deepfake for FLV. Specifically, 
RQ1 -- How vulnerable is FLV to deepfake-powered attacks? RQ2 -- How do the threats vary with concrete deepfake techniques? RQ3 -- What are the key factors impacting the attack effectiveness? RQ4 -- How would the practitioners improve the security of FLV? The answers to the above key questions are crucial for the deployment and use of FLV in practical settings.

\textbf{Our Work.} To answer these questions, in this paper, we design and implement \texttt{LiveBugger}, a framework that integrates various SOTA deepfake techniques for evaluating the security risks of FLV in a real-world setting.
Leveraging \texttt{LiveBugger}, we evaluate representative commercial FLV APIs provided by leading PaaS vendors. 
Then, we conduct an in-depth exploration of the factors impacting the attack effectiveness and conduct proof-of-concept attacks on real applications to further assess the threats in more real-world settings. Finally, we make a discussion of why the proposed deepfake-powered attack can break FLV, and provide suggestions to improve its security. 
We have reported our findings to the corresponding vendors and received their acknowledgment. In summary, we have made the following contributions.



\vspace{1pt}
{\em Framework ---} 
    We present \texttt{LiveBugger}, the first framework designed specifically to serve as a security evaluation framework for FLV in the deepfake era. At a high level, \texttt{LiveBugger} consists of three key components, as illustrated in Figure \ref{figure:livebugger},  namely \textit{Intelligence Engine}, \textit{Deepfaker Engine}, and \textit{Analysis Engine}.

1) \textit{Intelligence Engine}, which provides a complete set of authentication features supported by leading FLV PaaS vendors as well as a configurable interface to incorporate new vendors. The intelligence engine is able to automatically validate the claimed defense features using a customizable probing dataset. For example, BD\footnote{To minimize the ethical concern, we have replaced the vendor names with cryptonyms in this paper.} (one of the vendors that tops China's AI cloud services market) claims that its voice-based FLV API supports lip language detection. However, our analysis reveals that a video without any lip movements can also bypass this API. Overall, the intelligence engine facilitates efficient and fine-grained evaluation.


2) \textit{Deepfaker Engine}, which currently integrates six SOTA deepfake techniques. Based on the collected intelligence, the engine can synthesize the required fake videos for bypassing FLV effectively. For example, for the vendor without coherence detection, it concatenates pre-recorded videos satisfying the required actions as a driving video to synthesize the fake video for bypassing a target FLV. With a modular design, new deepfake techniques can be readily integrated into the engine.


3) {\em Analysis Engine}, which includes a set of information-rich, customizable metrics to support fine-grained evaluation of FLV, including liveness evasion rate, anti-deepfake evasion rate, face matching rate, and overall evasion rate.

\vspace{1pt}
{\em Evaluation ---} Leveraging \texttt{LiveBugger}, we conduct a systematic study of the most representative FLV APIs, including Image-, Silence-, Voice-, and action-based FLV. We make a number of interesting observations with the following highlights: 1) most vendors do not consider anti-deepfake detection in their FLV APIs, which are thus vulnerable to deepfake and threaten thousands of downstream applications; 2) even for the very few vendors which deploy anti-deepfake detection, the defense performance is problematic (e.g., while effective for videos of high visual quality, it fails to detect some poorly synthesized videos);  3) the security gain of the random process (e.g., random voice code or action sequence) in current voice-based FLV and action-based FLV is marginal.
Besides, we conduct proof-of-concept attacks on real applications to illustrate the practical implication brought by deepfake. The attacks show that most evaluated downstream  APPs (i.e., the clients of FLV APIs) are vulnerable to deepfake, thus threatening the security of millions of users. Our evaluation raises severe concerns about the commercial FLV APIs provided by PaaS vendors.

\vspace{1pt}
{\em Exploration ---} We further explore the impacting factors for the attack effectiveness, leading to a number of interesting findings: 1) the target image has more influence on the face reenactment methods for bypassing FLV, while the driving video has more influence on the face-swapping methods; 2) the adversary may exploit the bias (e.g., gender or race) in FLV to select the victim; and 3) adversarial training may benefit bypassing FLV. Based on such findings, we propose a customized two-stage method that improves the attack success rate of bypassing FLV by up to 70\%. 

\vspace{1pt}
{\em Security Suggestion ---} Based on our findings, we first discuss why the deepfake-powered attack can break FLV via comparing it with the presentation attack. Then, we provide suggestions for improving the security of FLV. For instance, the random code in voice-based FLV should not be limited to digits, but should be diversified to enhance the protection; action-based FLV should adopt actions that are difficult to synthesize for deepfake. We have reported our findings to affected vendors and received their acknowledgments. In response, one vendor has announced its engagement in a deepfake detection project to address this new threat.

We envision that our suggestions will shed light on developing more effective and robust FLV schemes in general.

\section{Background}

\subsection{Facial Liveness Verification}
A general overview of FLV is presented in Figure \ref{figure:online}. Below, we give a detailed introduction to the process of FLV, which mainly includes three steps. 

\textbf{Step 1.} A user interacts with the application and uses it to record his/her facial image/video.

\textbf{Step 2.} After collecting a user's facial media, the application will call the target FLV API with the recorded media.

\textbf{Step 3.} The API will verify the user's identity by analyzing the uploaded media. During the verification, the API first conducts the liveness detection, which is mainly used to verify whether the voice or action requirements are met and defend the presentation attack. After passing the liveness detection,  the API may further conduct deepfake spoofing detection if applicable. Finally, the API will perform face matching between the uploaded face and the reference face to verify the identity. The video/image that passes all the processes will be reported as a valid one.

According to the recorded media, existing FLV can mainly be divided into four categories: 1) \textbf{Image-based FLV}:  it performs liveness detection based on a static facial image uploaded by the user and mainly focuses on detecting the presentation attack; 2) \textbf{Silence-based FLV}: it performs liveness detection based on a facial video clip submitted by the user; 3) \textbf{Voice-based FLV}: the user is requested to speak the given digits while recording the facial video, while the FLV performs liveness detection by analyzing both the visual and audio signals; 4) \textbf{Action-based FLV}: the user is requested to act according to the given action sequence while recording the facial video, while the FLV performs liveness detection by checking whether the action requirements are met.


\subsection{Threat Model}

\label{threat_model}

To make our evaluation more practical, we focus on evaluating the security of FLV APIs provided by popular cloud vendors. Therefore, our evaluation is conducted under the black-box setting where an adversary cannot obtain any internal knowledge of the target API, like the liveness verification model, face matching model, etc. In this paper,
we mainly consider the one-shot setting --- the adversary can obtain one facial image of the victim since it is the lowest requirement for the adversary to bypass an FLV system. Therefore, it can approximately expose an API's worst-case vulnerability.

\begin{figure}[t]
	\centering
	
	\begin{minipage}[t]{0.49\columnwidth}
		\centering
		\includegraphics[width=\columnwidth]{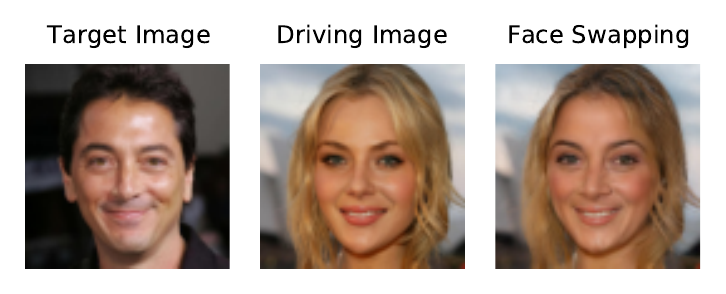}\\
		\label{image_faceswap}
	\end{minipage}
	\begin{minipage}[t]{0.49\columnwidth}
		\centering
		\includegraphics[width=\columnwidth]{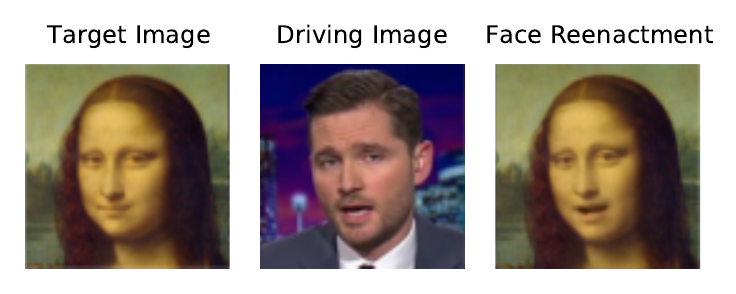}\\
		\label{image_facereen}	
	\end{minipage}%
	\centering
	\caption{\footnotesize Face swapping and reenactment. 
	}
	
	\label{fig:swap_reen}
\end{figure}

\subsection{Deepfake}
\label{sec_deepfake}

For studying the new threat brought by deepfake, we use SOTA deepfake techniques to evaluate the security risks of FLV.
In general, there are two types of deepfake techniques to synthesize fake images/videos: \textit{face swapping} and \textit{face reenactment} \cite{nirkin2019fsgan}. Both are able to synthesize the required video/image with respect to the given target image and the driving image/video, in which the target image provides the identity information, while the driving image/video provides the background/texture information (face-swapping) or motion information (face reenactment). 

As shown in the left plot of Figure \ref{fig:swap_reen}, face swapping transfers the identity information from the target image to the driving image/video. The driving video can be any video that satisfies the move/voice requirements (e.g., the adversary may use his/her own). 
In comparison, as shown in the right plot of Figure \ref{fig:swap_reen}, face reenactment uses the facial movement/expression deformation of the driving image/video to reenact the target image. 

To understand the new threat comprehensively, we will use both face-swapping and face reenactment to evaluate the security risks of FLV.

\section{Framework}

\begin{figure}[t]
	\centerline{\includegraphics[width=1.0\columnwidth]{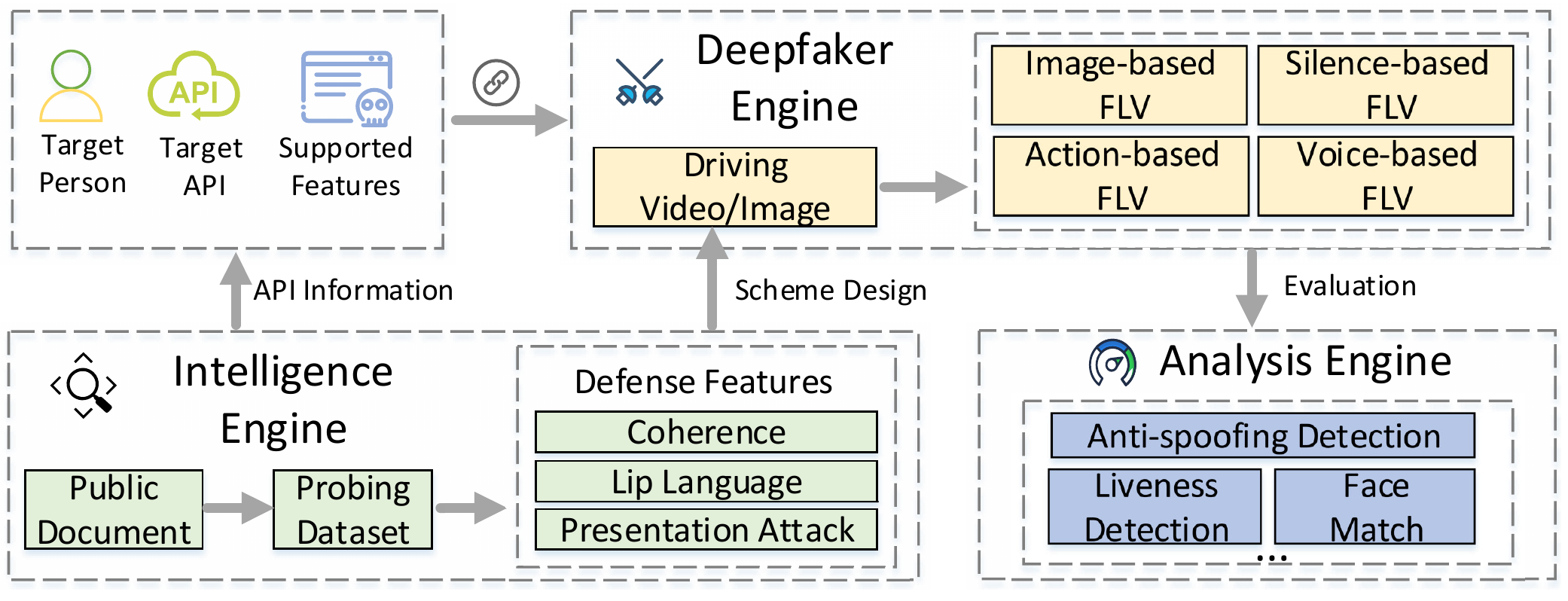}}
	\caption{\footnotesize Overview of \texttt{LiveBugger}.}	
	\label{figure:livebugger}
\end{figure}

To systematically evaluate the security risks of FLV APIs, we design and implement \texttt{LiveBugger}, an evaluation framework with high expandability. \texttt{LiveBugger} consists of three main components as illustrated in Figure \ref{figure:livebugger}:   \textit{Intelligence Engine}, \textit{Deepfaker Engine}, and \textit{Analysis Engine}. Below, we will introduce them in more detail.

\subsection{Intelligence Engine}

\textit{Intelligence Engine} is mainly used to construct a complete set of authentication features supported by the leading FLV PaaS vendors. Specifically, \textit{Intelligence Engine} collects the information from the public API documents provided by the vendors, which includes the action types, action sequence length range, the deployment of anti-deepfake detection, etc. Due to the marginal difference of the supported features provided by different vendors, we adopt the union set of them as the features that can be configured by \textit{Intelligence Engine}, which enables a configuration-based intelligence interface for new vendors to be evaluated. 
\textit{Intelligence Engine} currently has built-in configurations for six representative FLV vendors (the details of these vendors are introduced in Section \ref{evaluation}).

However, for security concerns, some implicit information  cannot be obtained from the official public information of vendors, e.g., the deployment of coherence detection. Additionally, 
the vendors may not support the claimed features in practice. To this end, we build a probing dataset inside the \textit{Intelligence Engine} to collect the implicit information and validate the claimed features. 
Specifically, \textit{Intelligence Engine} can automatically call the target API with the probing dataset, and then obtain the needed information based on the returned results. At present, it mainly uses the probing dataset to collect information on three defense features, including the deployment of  coherence detection, lip language detection, and presentation attack detection. The collected information is shown in Table \ref{api_intelligence}. Next, we introduce them in more detail.

\paragraphbe{Coherence Detection} Coherence detection checks whether the consecutive frames of a video are visually continuous.
To check if a target API deploys coherence detection, \textit{Intelligence Engine} includes a probing dataset consisting of a normal dataset and a corresponding disturbed one. Specifically, it uses several randomly selected facial videos to construct a normal dataset. Then, it scrambles the order of the frames in each selected video to get the corresponding disturbed dataset.
If the normal dataset and the disturbed one achieve similar success rates, then the target cloud vendor has not deployed the coherence detection; otherwise is the opposite.

\paragraphbe{Lip Language Detection} Lip language detection is to detect whether the lip movement in a video matches the corresponding audio signal. \textit{Intelligence Engine} includes three probing datasets for this detection: 1) a normal dataset containing videos whose audio signal matches the lip movement; 2) one perturbed dataset consisting of videos whose audio signal does not match the lip movement; 3) another perturbed dataset that includes videos with only audio signals but without any lip movement. If the bypass rate of the latter two datasets is much lower than that of the normal dataset, then the target cloud vendor has deployed the lip language detection; otherwise is the opposite. Besides, by comparing the bypass rate of the latter two datasets, \textit{Intelligence Engine} can check the level of lip language detection deployed by a cloud vendor.



\paragraphbe{Presentation Attack Detection} Similar to previous methods, \textit{Intelligence Engine} uses several randomly selected videos to construct two probing datasets, including a normal dataset and a replayed one, to check the deployment of presentation attack detection. Specifically, if the bypass rate of the replayed dataset is much lower than that of the normal dataset, then the presentation attack detection has been deployed by the target vendor; otherwise is the opposite.

\subsection{Deepfaker Engine} 
\label{deepfaker}

Leveraging \textit{Intelligence Engine}, the configuration information for a target API can be specified, which is then used by \textit{Deepfaker Engine} to synthesize the fake videos/images to evaluate the target API automatically. Specifically, \textit{Deepfaker Engine} incorporates several SOTA deepfake techniques that can work well in the one-shot setting.
Below, we briefly introduce the workflow of synthesizing the images/videos for bypassing different types of FLV API and defer the implementation details to Section \ref{deepfake_implementation}.

\paragraphbe{Image-based FLV} Many target images are unable to pass image-based FLV due to their background information (e.g., brightness and posture).  To this end,  this module takes several images that can pass image-based FLV as the driving images. Then, since face reenactment methods cannot change the background information, \texttt{LiveBugger} utilizes SOTA face-swapping methods to replace the background information of the target image with that of the driving image (i.e., replacing the identity of the driving image with that of the target image) for bypassing image-based FLV.


\paragraphbe{Silence-based FLV} It takes some randomly selected videos as the driving videos. Then, along with the target image, it uses SOTA  face swapping and reenactment methods to synthesize the fake videos for bypassing silence-based FLV.

\paragraphbe{Voice-based FLV} 
From the results returned by \textit{Intelligence Engine} (see the details in Table \ref{api_intelligence}), we find that most evaluated vendors have not deployed lip language detection in their voice-based FLV APIs. 
Therefore, we can directly import the required audio signal to the synthesized video to evaluate voice-based FLV APIs. Specifically, this module synthesizes fake videos based on the target image and a randomly selected driving video with lip movement (some APIs only detect lip movement). Then, after receiving the random digits, it uses a voice synthesis model, which can be the voice synthesis API provided by cloud vendors, to synthesize the required audio signal and import it to the synthesized video. For the few APIs that deploy lip language detection, one needs to record a driving video with the matched lip movement interactively.

\paragraphbe{Action-based FLV}
From Table \ref{api_intelligence},  we find that all the evaluated APIs have not deployed the coherence detection. Therefore, the driving video can be prepared by directly stitching the pre-recorded videos of the required actions. Accordingly, this module provides built-in videos of different actions from volunteers. 
Based on the stitched driving video and the target image, it synthesizes the corresponding fake video for bypassing action-based FLV.
At the same time, we notice that a few demo APPs evaluated in Section \ref{global} use the coherence detection (see details in Section \ref{global}). For evaluating them, after receiving the action sequence, one needs to record a video as the driving video since its natural coherence.

\begin{figure*}[ht]
	\centerline{\includegraphics[width=0.9\textwidth]{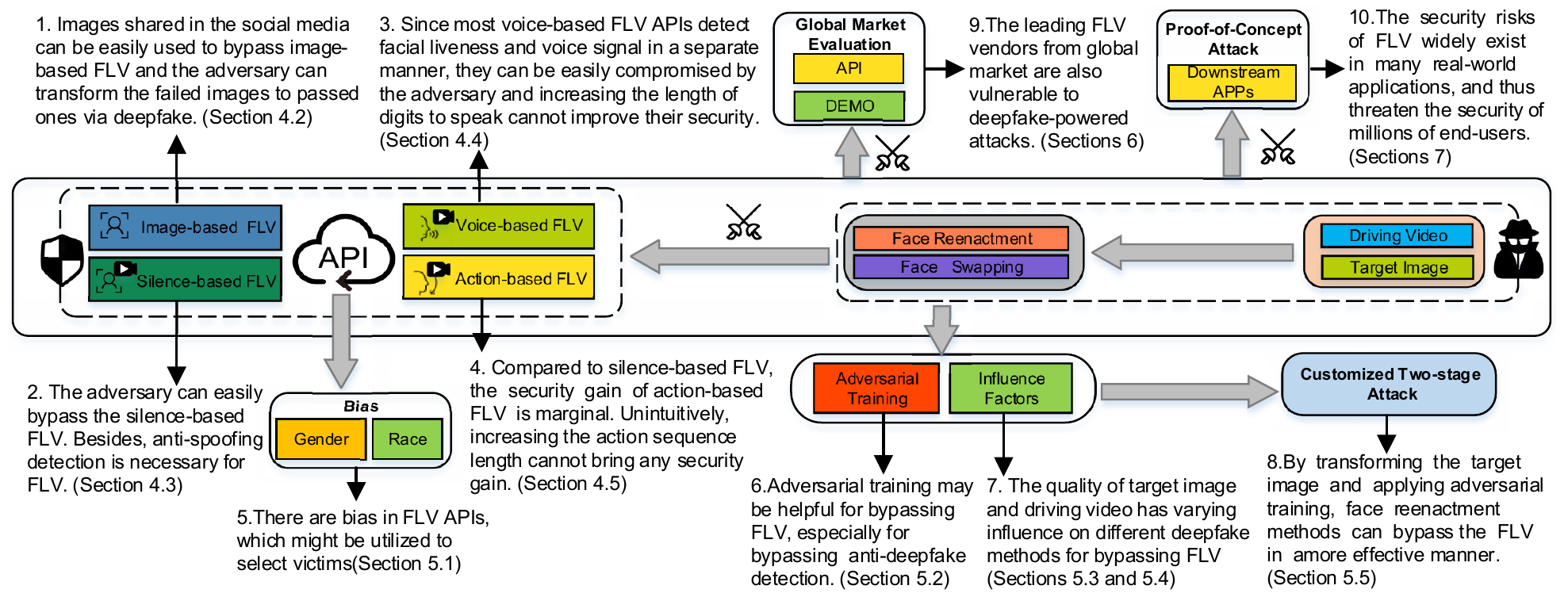}}
	\caption{\footnotesize Overview of the insight in our work. Remarks 1-4 denote the evaluation insights, and Remarks 5-10 denote the exploration insights.}	
	\label{figure:overview}
\end{figure*}

\subsection{Analysis Engine} 
Different vendors provide FLV in various forms. For flexibility, some vendors separate face matching from FLV  and offer it as an independent API. When conducting verification, FLV often returns a frame for testing (test frame). The \textit{Analysis Engine} uses the test frame and a facial image of the target individual (reference image) to call the corresponding face matching API to perform verification. 
For a few APIs which do not return the 
test frame, we randomly sample frames to conduct face matching and report average results. For several other vendors, face matching is integrated into their FLV APIs, which return both liveness and face matching results. To quantitatively characterize the threats, we propose the following metrics.

\vspace{1pt}
\noindent\textbf{Liveness Evasion Rate.} It  measures the rate of images/videos that meet the action/voice requirements (if applicable) and pass the presentation attack detection. A higher liveness evasion rate implies lower security of the FLV.

\vspace{1pt}
\noindent\textbf{Anti-deepfake Evasion Rate.} Certain cloud vendors deploy anti-deepfake detection mechanisms. The anti-deepfake detection results are returned to the users separately. Therefore, we use anti-deepfake evasion rate to measure the rate of synthesized images/videos that evade the anti-deepfake detection. A larger evasion rate implies higher attack effectiveness. 

\vspace{1pt}
\noindent\textbf{Face Matching Rate.} It measures the rate of synthesized media that pass the face matching mechanism. A larger matching rate implies better quality of the synthesized media.

\vspace{1pt}
\noindent\textbf{Overall Evasion Rate.} It assesses the overall security of the target API by measuring the fraction of  synthesized media that evade liveness detection, deepfake spoofing detection, and face matching simultaneously. A larger rate implies higher attack effectiveness or less security of the target API.

These metrics allow us to characterize the threats from various defense perspectives (e.g., liveness detection, deepfake detection, face matching) and in a fine-grained manner, leading to a set of interesting findings.

\begin{table*}[!ht]{\footnotesize
	\setlength{\tabcolsep}{2pt}
		\centering
\begin{threeparttable}		
 \begin{tabular}{cccccccccccccccc}

\multirow{5}{*}{Platform} & \multicolumn{15}{c}{Liveness Type}                                                                                                                                                                                                                                                                                                                                                                                                                                                                                                                                                                                                                                                                                                                                                                                                                                    \\ \cline{2-16} 
                          & \multirow{4}{*}{Image} & \multirow{4}{*}{} & \multicolumn{13}{c}{Video}                                                                                                                                                                                                                                                                                                                                                                                                                                                                                                                                                                                                                                                                                                                                                                                                                             \\ \cline{4-16} 
                          &                        &                   & \multicolumn{9}{c}{Type}                                                                                                                                                                                                                                                                                                                                                                                                                                                                                                                                           & \multirow{3}{*}{} & \multicolumn{3}{c}{Common Detection}                                                                                                                                                                                                                          \\ \cline{4-12} \cline{14-16} 
                          &                        &                   & \multirow{2}{*}{Silence} & \multicolumn{4}{c}{Voice}                                                                                                                                                                                                                                                  &  & \multicolumn{3}{c}{Action}                                                                                                                                                                                                                              &                   & \multirow{2}{*}{\begin{tabular}[c]{@{}c@{}}Anti-\\ deepfake\\ Detection\end{tabular}} & \multirow{2}{*}{\begin{tabular}[c]{@{}c@{}}Coherence\\ Detection\end{tabular}} & \multirow{2}{*}{\begin{tabular}[c]{@{}c@{}}Replay\\ Attack\\ Detection\end{tabular}} \\ \cline{5-8} \cline{10-12}
                          &                        &                   &                          & \begin{tabular}[c]{@{}c@{}}Voice Cide \\ Length\\ Range\end{tabular} & \begin{tabular}[c]{@{}c@{}}Voice\\ Code\\ Type\end{tabular} & \begin{tabular}[c]{@{}c@{}}Default\\ Code\\ Length\end{tabular} & \begin{tabular}[c]{@{}c@{}}Lip\\ Language \\ Detection\end{tabular} &  & \begin{tabular}[c]{@{}c@{}}Action\\ Length\\ Range\end{tabular} & \begin{tabular}[c]{@{}c@{}}Default\\ Action\\ Length\end{tabular} & \begin{tabular}[c]{@{}c@{}}Action\\ Type\end{tabular}                                                             &                   &                                                                                       &                                                                                &                                                                                      \\ \hline
BD                     & $\fullcirc$            &                   & $\fullcirc$              & 3 - 6                                                                & Digits                                                      & 3 - 6                                                           & $\emptycirc$                                                        &  & 1 - 3                                                           & 1 - 3                                                             & \begin{tabular}[c]{@{}c@{}}Blink, Turn Right\\ Turn Left, Look Up\\ Chin Down,\\ Turn Right and Left\end{tabular} &                   & $\fullcirc$                                                                           & $\emptycirc$                                                                   & $\fullcirc$                                                                          \\
TC                   & $\fullcirc$            &                   & $\fullcirc$              & 1 - 6                                                                & Digits                                                      & 4                                                               & $\halfcirc$                                                         &  & 1 - 2                                                           & 2                                                                 & \begin{tabular}[c]{@{}c@{}}Blink,\\ Open Mouth\end{tabular}                                                       &                   & $\fullcirc$                                                                           & $\emptycirc$                                                                   & $\fullcirc$                                                                          \\
HW                    & $\fullcirc$            &                   & $\emptycirc$             & \multicolumn{4}{c}{$\emptycirc$}                                                                                                                                                                                                                                           &  & 1 - 4                                                           & 1                                                                 & \begin{tabular}[c]{@{}c@{}}Turn Left, Turn Right,\\ Blink, Open Mouth\end{tabular}                                &                   & $\emptycirc$                                                                          & $\emptycirc$                                                                   & $\fullcirc$                                                                          \\
CW                 & $\fullcirc$            &                   & $\fullcirc$              & 4 - 6                                                                & Digits                                                      & 4 - 6                                                           & $\fullcirc$                                                         &  & \multicolumn{3}{c}{$\emptycirc$}                                                                                                                                                                                                                        &                   & $\emptycirc$                                                                          & $\emptycirc$                                                                   & $\fullcirc$                                                                          \\
ST                 & $\emptycirc$           &                   & $\fullcirc$              & 4                                                                    & Digits                                                      & 4                                                               & $\emptycirc$                                                        &  & \multicolumn{3}{c}{$\emptycirc$}                                                                                                                                                                                                                        &                   & $\emptycirc$                                                                          & $\emptycirc$                                                                   & $\fullcirc$                                                                          \\
iFT                  & $\fullcirc$            &                   & $\fullcirc$              & \multicolumn{4}{c}{$\emptycirc$}                                                                                                                                                                                                                                           &  & \multicolumn{3}{c}{$\emptycirc$}                                                                                                                                                                                                                        &                   & $\emptycirc$                                                                          & $\emptycirc$                                                                   & $\fullcirc$                                                                          \\ 
\end{tabular}
	
	\caption{\footnotesize API intelligence collected from cloud platforms. $\fullcirc$ denotes full support; $\halfcirc$ denotes partial support; $\emptycirc$ denotes no support.}
	\label{api_intelligence}
	
\end{threeparttable}}
	
\end{table*}


\section{Evaluation}
\label{evaluation}

In this section, we first introduce the overall experimental setting, including the vendors, datasets, and deepfake methods. Then, leveraging \texttt{LiveBugger}, we systematically evaluate the FLV APIs provided by the leading FLV PaaS vendors.

\subsection{Overall Experimental Setting}

\paragraphbe{Target Vendors} 
To make the evaluation more practical, we leverage \texttt{LiveBugger} to evaluate the FLV APIs provided by popular commercial cloud vendors according to the facial recognition market share \cite{market3}. 
Specifically, we evaluate the six most representative FLV vendors, including BD, TC, HW, CW, ST, and iFT (to minimize the ethical concern, we have replaced the vendor names with cryptonyms). The reasons behind considering these vendors are as follows. 1) \textbf{BD} and \textbf{TC} are one of the vendors with the largest China's AI cloud services market and the greatest number of face-related API calls, respectively; 2) \textbf{HW} is one of the vendors with the largest China's public cloud market; 3) \textbf{CW} is one of the vendors with the fastest growth rate in computer vision and is becoming the leader; 4) \textbf{ST} is one of the largest computer vision vendor; 5) \textbf{iFT} is one of the vendors with the largest China's AI software market.
\texttt{LiveBugger} collects the configurations, as shown in Table \ref{api_intelligence}, for the supported authentication features of the FLV APIs provided by these vendors. Table \ref{api_intelligence} shows the supported authentication features for each target vendor such as voice code length range and supported the action, which facilitates an automated evaluation. Given an evaluation configuration,   \texttt{LiveBugger} automatically evaluates the target APIs using the target images and the synthesized images/videos.
To better illustrate the threat surface, we also evaluate some representative FLV APIs from the global market in Section \ref{global}.

\paragraphbe{Datasets}  
First, our evaluation needs an image dataset to provide the target images for deepfake synthesis and the reference images for face matching. Therefore, we use  the live images from  CelebA-Spoof \cite{zhang2020celeba} as \textbf{the image dataset}. CelebA-Spoof is a face anti-spoofing dataset that has 625,537 images crawled from social media, which includes 43 rich attributes on the face, environment, and spoof types.

Second, our study needs a video dataset to provide  driving videos. Therefore, we use the live videos from SiW-M \cite{liu2019deep} as \textbf{the video dataset}. SiW-M provides live and spoof (e.g., replay) videos from 165 subjects \cite{liu2019deep}.



 

\paragraphbe{Deepfake Methods and Implementation} 
\label{deepfake_implementation}
According to the threat model, the used deepfake method to evaluate an FLV API should meet the following requirements: 1) it should be identity-agnostic, i.e., it does not need to train a new model for a new target person; 2) it can synthesize the required video based on one facial image of the target person; 3) when synthesizing videos/images, its latency needs to be acceptable; otherwise, a timeout of the target FLV API will occur. Therefore, \texttt{LiveBugger} incorporates six SOTA deepfake methods that meet the above requirements, including X2Face \cite{wiles2018x2face}, ICface \cite{tripathy2020icface},  FSGAN$_S$  \cite{nirkin2019fsgan}, FSGAN$_R$ \cite{nirkin2019fsgan}, First Order Method Model (FOMM)  \cite{siarohin2019first} and 
FaceShifter  \cite{li2019faceshifter}.  We present their details, like technical highlights and categories, in Appendix \ref{appendix_deepfake}. Note that, except for FaceShifter, we use the open-source code published by the authors. Since FaceShifter is not open-source, we reproduce it according to our understanding of the original paper.  All of our experiments are conducted on a server with two Intel Xeon E5-2640 v4 CPUs running at 2.40GHz, 256 GB memory, 4TB HDD, and 4 GeForce GTX 1080 Ti GPU cards.

Before diving into the detailed evaluation results, we present an overview of the core insights (Remarks 1 to 4) in Figure \ref{figure:overview}. For each vendor, we evaluate four types of FLV APIs if available, including image-based FLV, silence-based FLV, voice-based FLV, and action-based FLV, the insights of which correspond to Remarks 1 - 4, respectively.



\subsection{Image-based FLV }
\label{eval_iflv}

 \begin{table}[t]{\footnotesize
\setlength{\tabcolsep}{3pt}
	\centering
	
\begin{tabular}{ccccc}
Platform  & \begin{tabular}[c]{@{}c@{}}Liveness \\ Evasion\end{tabular} & \begin{tabular}[c]{@{}c@{}}Anti-deepfake \\ Evasion\end{tabular} & \begin{tabular}[c]{@{}c@{}}Face \\ Matching\end{tabular} & \begin{tabular}[c]{@{}c@{}}Overall \\ Evasion\end{tabular} \\ \hline
BD    & 75\%                                                              & \textbf{90\%}                                         & 99\%                                                           & 68\%                                                             \\
TC   & 53\%                                                              & 85\%                                                                   & \textbf{100\%}                                & 42\%                                                             \\
HW    & 70\%                                                              & -                                                                      & 99\%                                                           & 70\%                                                             \\
CW & 97\%                                                              & -                                                                      & \textbf{100\%}                               & 97\%                                                             \\
iFT  & \textbf{99\%}                                    & -                                                                      & \textbf{100\%}                                & \textbf{99\%}                                \\ 
\end{tabular}
	\caption{\footnotesize Evaluation of legitimate images against FLV (false positive rate = 100\% - evasion/matching rate).}
	\label{image_risk}}
\end{table}

Recall that image-based FLV performs liveness detection based on the uploaded static image and focuses on detecting the presentation attack. To assess the performance of a given API, we first measure its false-positive rate (FPR) using 200 legitimate images sampled from the image dataset, with results presented in Table \ref{image_risk}. A lower overall evasion rate implies higher FPR. Observe that although the live image may be directly used to evade image-based FLV, many of them fail to pass the target API due to the background information (e.g., brightness and posture). Below, we consider an adversarial setting: the adversary attempts to transform the failed image into a successful one via deepfake techniques. 



\paragraphbe{Target Images and Driving Images}
For each vendor, we first sample 100 images that fail to pass the target image-based FLV API from the image dataset as the target images. Then, for each target image, we select another image with the same identity as the reference image for identity verification.  For the driving images, \texttt{LiveBugger} randomly selects 10 images with other identities that pass the target image-based FLV API from the image dataset.
Utilizing the face detector \cite{bulat2017far}, we crop every image  to 256$\times$256 pixels, and without explicitly specified, the video preprocessing in the rest of the paper is the same as that of the image.




\paragraphbe{Security Evaluation} Since ST does not provide image-based FLV, we evaluate the image-based FLV APIs from the other five vendors. For a given target API and each target image in the corresponding evaluation dataset, \texttt{LiveBugger} transforms the image using its \textit{Deepfaker Engine} and then uses the transformed one to evaluate the target API. Note that, as stated in Section \ref{deepfaker}, since face reenactment methods cannot swap the background information of the failed image, we focus on face-swapping methods (FaceShifter and FSGAN$_S$) in this section.

\begin{figure}[t]

	\centerline{\includegraphics[width=0.8\columnwidth]{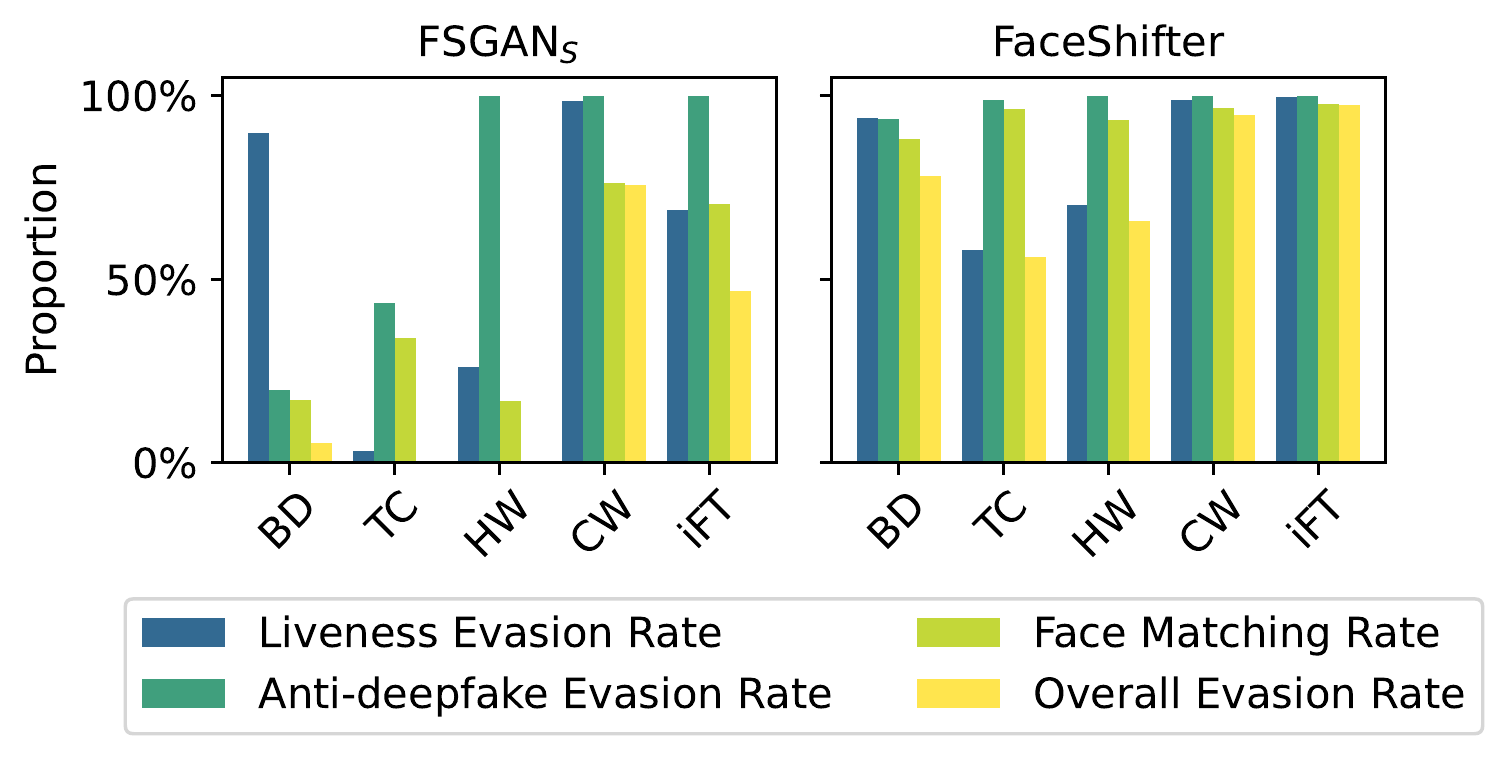}}
	\caption{\footnotesize Evaluation of transformed images against image-based FLV APIs (note: HW, CW, and iFT have not deployed anti-deepfake detection, their default anti-deepfake evasion rates are set as 100\%).}
	
	\label{fig:image_evaluation}
\end{figure}

The results are shown in Figure \ref{fig:image_evaluation}, and 
we have the following observations. 1) Image-based FLV systems are highly vulnerable to deepfake-powered attacks. For instance, FaceShifter achieves a 95\%+ overall evasion rate on CW and iFT. 
While for Vendor and TC, as shown in Figure \ref{fig:image_evaluation} and Table \ref{image_risk}, the synthesized images even achieve a higher overall evasion rate than the legitimate images (BD: 78\% vs. 68\%, TC: 56\% vs. 42\%). 
2) The anti-deepfake detection deployed by TC and BD is unreliable.  Specifically, FaceShifter achieves 94\% and 99\% anti-deepfake evasion rate on BD and TC, respectively, even higher than the legitimate images (BD: 94\% vs. 90\%, TC: 99\% vs. 85\%).
3) Combining with Table \ref{image_risk} (from which the FPR of each vendor can be derived), we observe that a target API with higher FPR often offers stronger security. For example, TC has higher FPR but also more robustness compared to other evaluated vendors. We speculate that this is due to the utility-security trade-off: 
FLV often uses a threshold to adjust this trade-off. The threshold may vary in different scenarios (e.g., different lighting conditions). We use the thresholds recommended by the target vendors in our evaluation.

\subsection{Silence-based FLV}
\label{sec:silent_liveness}

\begin{table}[t]{\footnotesize
\setlength{\tabcolsep}{2pt}
	\centering
\begin{tabular}{ccccc}
Platform  & \begin{tabular}[c]{@{}c@{}}Liveness\\ Evasion Rate\end{tabular} & \begin{tabular}[c]{@{}c@{}}Anti-deepfake\\ Evasion Rate\end{tabular} & \begin{tabular}[c]{@{}c@{}}Face\\ Matching Rate\end{tabular} & \begin{tabular}[c]{@{}c@{}}Overall\\ Evasion Rate\end{tabular} \\ \hline
BD     & 67\%                                                             & 37\%                                                                  & 100\%                                                         & 25\%                                                            \\
TC   & 72\%                                                             & 100\%                                                                 & 100\%                                                         & 72\%                                                            \\
ST & 62\%                                                             & -                                                                     & 99\%                                                          & 62\%                                                            \\
CW & 97\%                                                             & -                                                                     & 100\%                                                         & 97\%                                                            \\
iFT  & 62\%                                                             & -                                                                     & 100\%                                                         & 62\%                                                            \\ 
\end{tabular}
	\caption{\footnotesize Evaluation  using legitimate  videos to measure false positive rate. }
	\label{silent_normal}}
\end{table}

Silence-based FLV utilizes an uploaded video to verify the identity of a target person. It does not require any additional action, like speaking digits or head movements. Like image-based FLV, we first  evaluate the FPR of the target silence-based FLV APIs using randomly selected legitimate videos. Table \ref{silent_normal} shows that the FPR of silence-based FLV is much higher than the vendors' claims.
However, according to the threat model, we cannot obtain the video of a target person. Therefore, in this section, we want to answer the following question:  can an adversary utilize a victim's facial image to bypass the silence-based FLV via deepfake?




\paragraphbe{Driving Videos and Target Images} 
We randomly select 40 images from the image dataset as the target images. Similarly, for each target image, we select another image with the same identity as the reference image for identity verification. Besides, \texttt{LiveBugger} randomly selects five videos from the video dataset as the driving videos.

\begin{figure}[t]

	\centerline{\includegraphics[width=0.9\columnwidth]{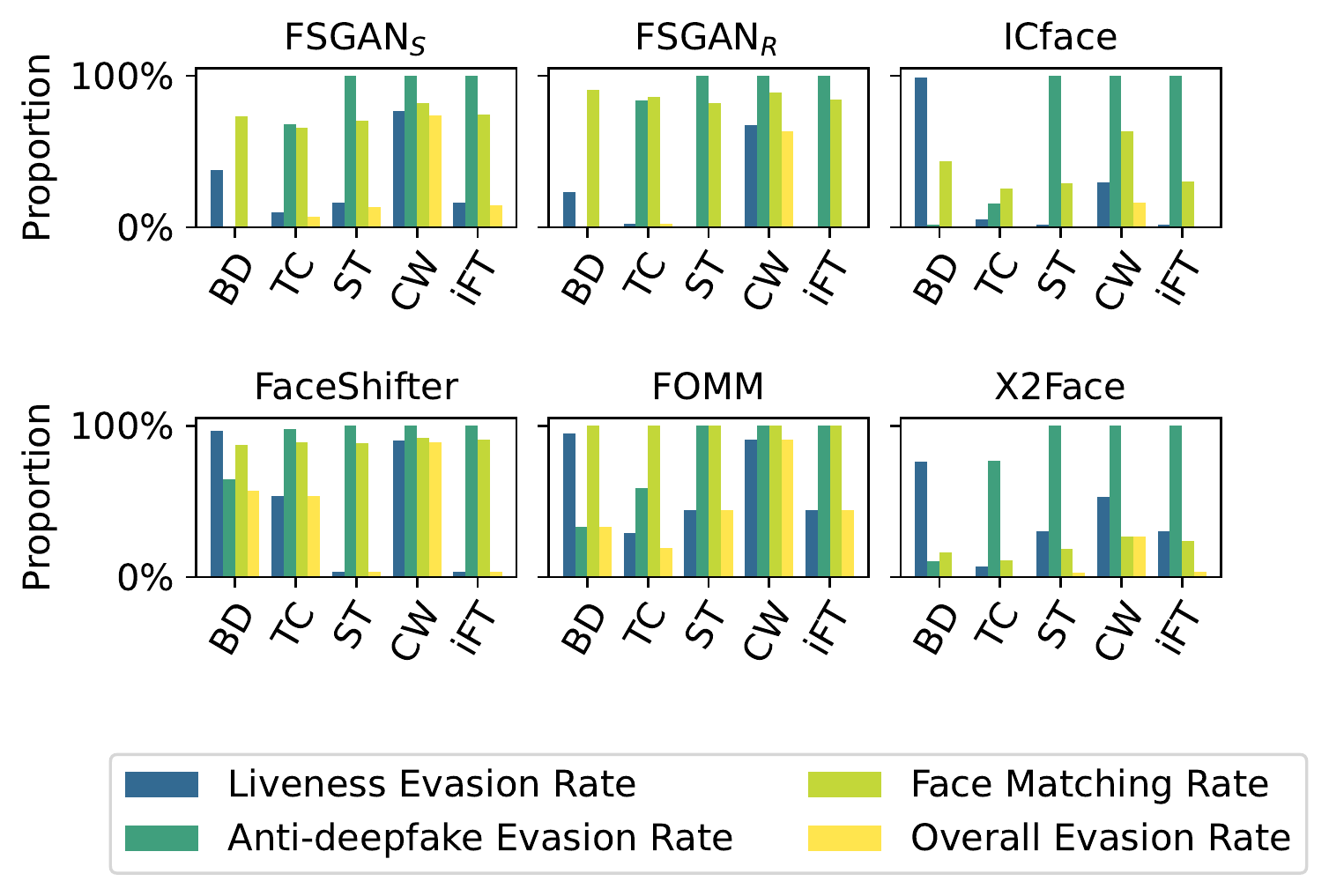}}
	\caption{\footnotesize Evaluation of silence-based FLV APIs. Since  ST, CW, and iFT have not deployed anti-deepfake detection, we assign 100\% to their anti-deepfake evasion rate.}
	
	\label{fig:silent_evaluation}
\end{figure}

\paragraphbe{Security Evaluation}
We utilize \texttt{LiveBugger} to synthesize fake videos based on the selected driving videos and target images and then evaluate the target API with the synthesized videos. Since HW does not provide silence-based FLV, we do not show its evaluation. Figure \ref{fig:silent_evaluation} shows the evaluation results of silence-based FLV. Note that certain deepfake methods (e.g., ICface) attain the overall evasion rate/liveness evasion rate of 0, which results in an invisible overall evasion rate/liveness evasion rate in the plots.
According to Figure \ref{fig:silent_evaluation}, we have the following observations. 1) An adversary can easily bypass the silence-based FLV API. The overall evasion rate on each platform can reach up to 40\%+. Especially, for CW, its overall evasion rate can reach up to 90\%+, which means that the silence-based FLV API of CW practically performs almost no function. 2) Anti-deepfake detection is necessary for liveness verification. For example, the results using ICface in Figure \ref{fig:silent_evaluation} show that although BD's liveness evasion rate is near 100\%, its overall evasion rate is 0 thanks to the deployment of anti-deepfake detection. The importance of deploying anti-deepfake detection is also confirmed by the results of FOMM, which show that although BD and CW have similar liveness evasion rate, BD has a much lower overall evasion rate than that of CW due to its better anti-deepfake detection ability.
3) The anti-deepfake detection deployed by a few vendors may be problematic. Figure \ref{fig:silent_evaluation} shows that compared to FaceShifter, FOMM achieves higher face matching rate but lower anti-deepfake evasion rate.  This is due to that it may successfully detect synthesized videos with high quality (i.e., high face matching rate) but fail to detect low-quality ones. For example,  Figure \ref{fig:extract_frame} in Appendix \ref{appendix_sflv} shows several frames extracted from a low-quality synthesized video and a high-quality one, respectively. In Figure \ref{fig:extract_frame}, each frame of the second row has high quality, while the corresponding video cannot bypass the anti-deepfake detection. However, the video consisting of the low-quality frames in the first row is able to bypass the detection.
Secondly, fake videos can achieve a much higher anti-deepfake evasion rate than real live videos. As shown in Figure \ref{fig:silent_evaluation} and Table \ref{silent_normal}, the videos synthesized by FaceShifter can achieve around 60\% anti-deepfake evasion rate on BD, while the real live videos only achieve 37\%. Therefore, anti-deepfake detection should be further improved.

\begin{figure}[t]
	\centerline{\includegraphics[width=0.6\columnwidth]{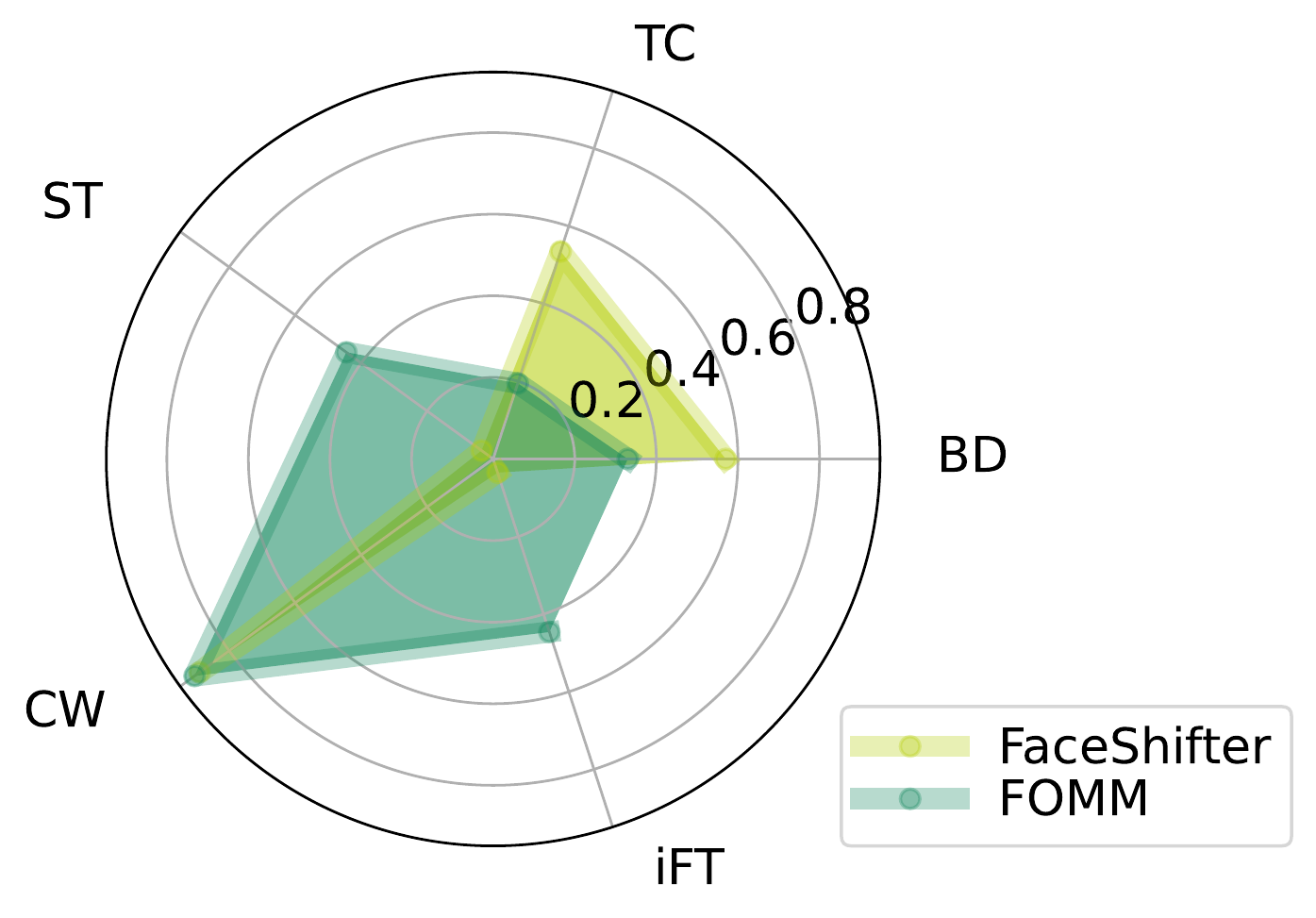}}
	\caption{\footnotesize Performance (overall evasion rate) comparison of FaceShifter and FOMM on different vendors. }	
	\label{figure:radar_compare}
\end{figure}

Compared to other methods, FaceShifter and FOMM  have a much higher overall evasion rate. For comparing them more clearly, we show their overall evasion rates on different vendors in Figure \ref{figure:radar_compare}. They both have a very high overall evasion rate on CW. For BD and TC, FaceShifter performs better, while FOMM performs better on ST and iFT. This indicates that different deepfake methods have different adaptability on different vendors. Therefore, a vendor should consider as many deepfake methods as possible to develop a more general and robust defense method.


	

%

\subsection{Voice-based FLV}
Voice-based FLV requires a user to speak given digits while recording the corresponding facial video to verify his/her identity.
Intuitively, since voice-based FLV introduces a random process based on silence-based FLV, it should largely mitigate the security risks.
Similar to Section \ref{sec:silent_liveness}, we use only one facial image to synthesize the required video for bypassing voice-based FLV. Here, the experiments focus on evaluating the security impact of the following key factors: 1) the random voice process, 2) the lip language detection, and 3) the digit length.


\paragraphbe{Driving Videos and Target Images} 
We keep the target images and reference images  the same as the images selected in Section \ref{sec:silent_liveness}. Then, \texttt{LiveBugger} randomly selects five videos with lip movements from the video dataset as the driving videos, since some vendors (e.g., TC) consider lip movements during the verification.


\begin{figure}[t]

	\centerline{\includegraphics[width=0.9\columnwidth]{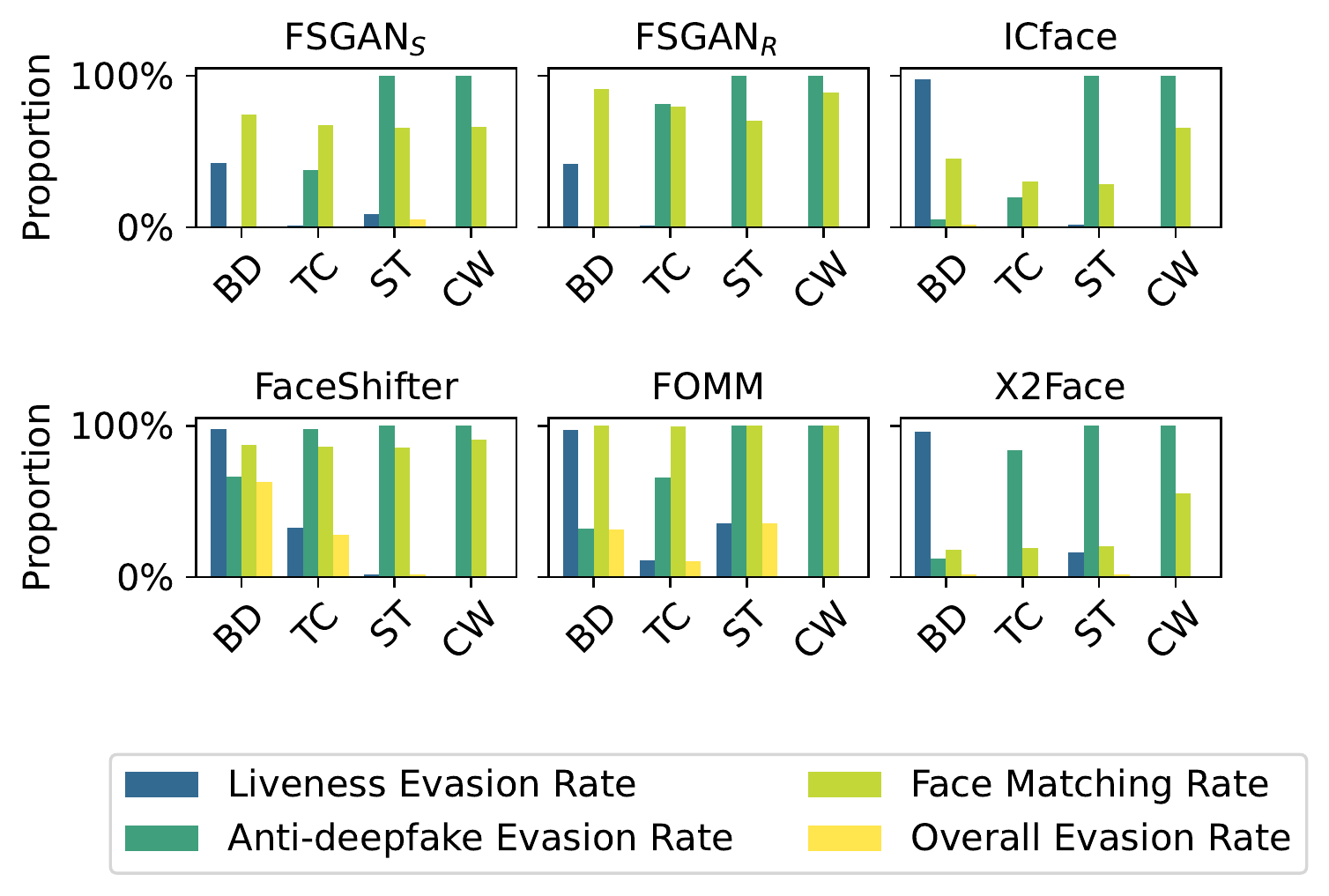}}
	\caption{\footnotesize Evaluation  of voice-based FLV APIs. Since ST and  CW have not deployed anti-deepfake detection, we assign 100\% to their anti-deepfake evasion rate.}
	
	\label{fig:voice_evaluation}
\end{figure}

\paragraphbe{Security Evaluation}
According to Table \ref{api_intelligence}, four vendors, including BD, TC, CW, and ST, provide voice-based FLV. Therefore, we will evaluate the security risks of the voice-based FLV APIs on these vendors. During the evaluation, \texttt{LiveBugger} utilizes the voice synthesis API from TC to synthesize the required audio signal. Note that, the choice of the voice synthesis API does not affect the evaluation result since the voice recognition process in voice-based FLV can correctly recognize the synthesized voice.

The evaluation results are shown in Figure \ref{fig:voice_evaluation}. Comparing the results of Figures \ref{fig:silent_evaluation} and \ref{fig:voice_evaluation}, we can see that except for CW,  voice-based FLV APIs can also be easily bypassed.  For example,  for FaceShifter in Figure \ref{fig:voice_evaluation}, the  overall evasion rate of BD can reach up to 60\%+, which is even higher than that in  silence-based FLV.  We speculate that this is because the target API detects the facial liveness and the audio signal separately. Specifically, the target API imports an independent speech recognition process on the basis of the silence-based FLV API to check whether the audio signal matches the given digits without considering lip language detection. Based on the returned API results from BD, we can observe that the audio signal perfectly matches the given digits. Thus, no security gain can be obtained based on the current implementation of voice-based FLV. Interestingly,  BD claims that their API supports lip language detection. However, we find that it is not valid. Even the video without any lip movement can bypass it. Similar security risks also exist in TC and ST. However, the voice-based FLV API of TC shows a slightly better defense performance than its silence-based FLV. The reason is that TC additionally detects lip movements but does not require the movements to match the given digits. Due to the imperfection of deepfake methods, the lip movements in some synthesized videos are not obvious, which results in a slight security improvement of the API. As for CW, since it deploys the lip language detection, we give a separate evaluation below.



\paragraphbe{Lip Language Detection} As shown in Figures \ref{fig:silent_evaluation} and \ref{fig:voice_evaluation},  since the driving video has unmatched lip movements with the given digits, the overall evasion rate of CW's voice-based FLV API decreases to 0. To this end, we use the customized driving videos with matched lip movements to evaluate CW and show the results in Figure \ref{figure:voice_customized_video} of Appendix \ref{appendix_vflv}. We find that  even though the voice-based FLV API deploys lip language detection, it still has high-security risks.

\paragraphbe{Length of Given Digits} Intuitively, because of the trade-off between security and utility, increasing the length of the given digits (decreasing the utility) should improve the security of voice-based FLV. However, the evaluation results show that increasing the length of the digits at the cost of utility does not improve such security. Due to the space limitations, we place the details in Appendix \ref{appendix_vflv}.

\subsection{Action-based FLV}
Similar to voice-based FLV, action-based FLV also introduces a random process based on silence-based FLV. The difference is that it requires a person to make head movements according to the given action sequence when recording the corresponding facial video. According to the threat model, we also utilize one single facial image of the victim to evaluate the security of action-based FLV. In this section, we mainly answer the following key questions: 1) does the random action sequence improve the security of silence-based FLV? 2) is there any security difference between different actions? 3) does the action sequence length affect the security of an action-based FLV API?

\begin{figure}[t]

	\centerline{\includegraphics[width=0.8\columnwidth]{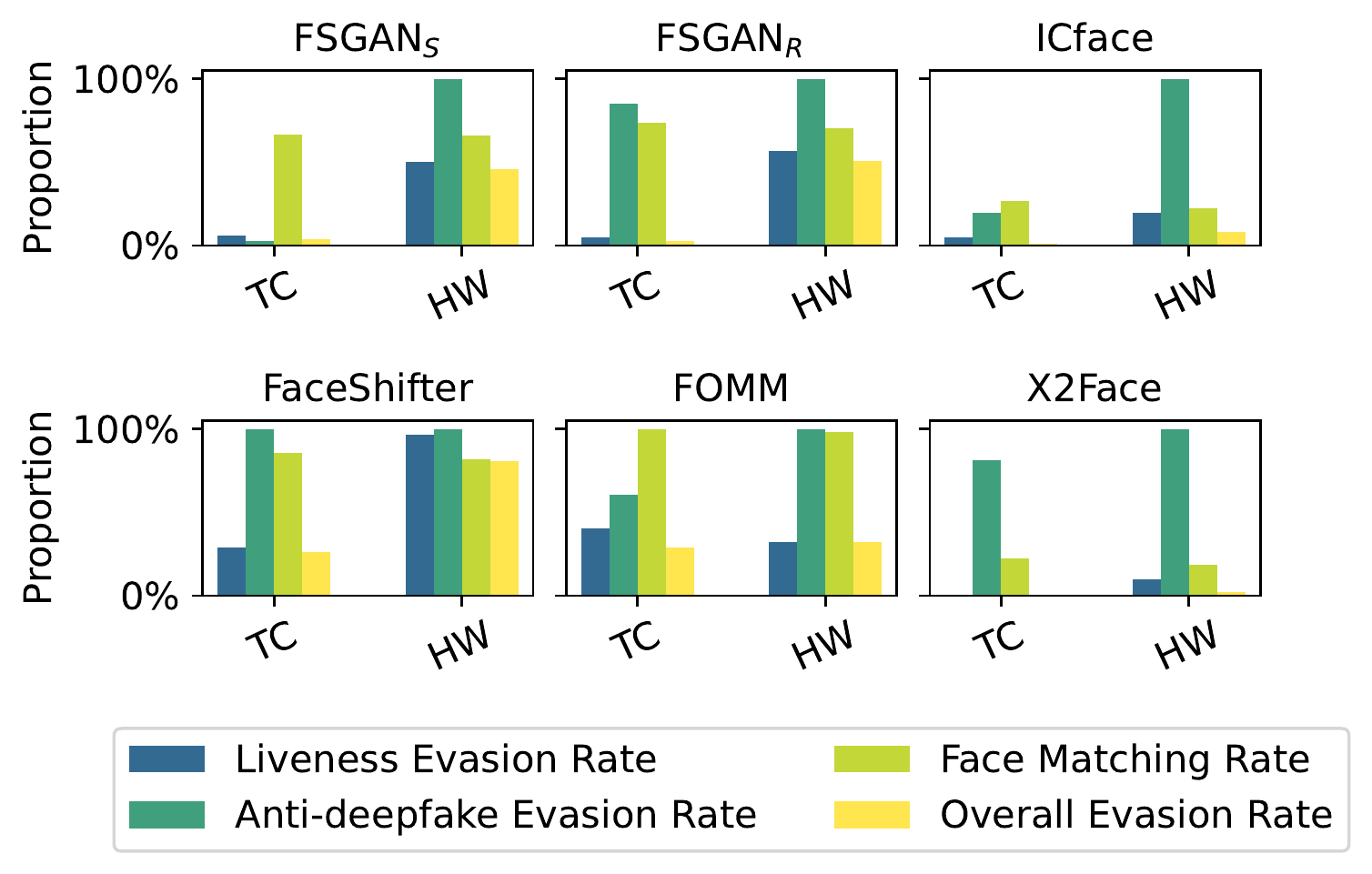}}
	\caption{\footnotesize Evaluation  of action-based FLV APIs. Since HW has not deployed the anti-deepfake detection, we assign 100\% to its anti-deepfake evasion rate.}
	
	\label{fig:action_evaluation}
\end{figure}

\paragraphbe{Driving Videos and Target Images}
We also keep the target and reference images the same as those in silence-based FLV. Additionally, we recruit five volunteers to record videos for each of the actions supported by the vendors. Then, \texttt{LiveBugger} can stitch the videos of the required actions to construct the driving video.


\paragraphbe{Security Evaluation} 
According to Table \ref{api_intelligence}, three vendors, including BD, TC, and HW, provide  action-based FLV. However, 
we find that the action-based FLV API on BD has an implementation problem: a video with the required actions incurs a calling error of the action-based FLV API but works normally with the silence-based FLV API. After contacting customer service, it is confirmed to be a video encoding problem. Since it has not been resolved so far, we evaluate TC and HW in this section.

The evaluation results of  action-based FLV are shown in Figure \ref{fig:action_evaluation}. From Figures \ref{fig:silent_evaluation} and \ref{fig:action_evaluation}, we have the following observations. 1) Action-based FLV can be bypassed very easily. For example, as shown by the FaceShifter of Figure \ref{fig:action_evaluation}, the liveness evasion rate on HW can reach up to 97\%, and the overall evasion rate can reach up to 80\%, which brings serious risks to the downstream applications. 2) Compared to silence-based FLV, the security gain of action-based FLV is marginal. Especially, as shown by the FOMM results in Figures \ref{fig:silent_evaluation} and \ref{fig:action_evaluation}, the overall evasion rate of action-based FLV on TC is even slightly higher than that of silence-based FLV. 3) As action-based FLV requires large movements like looking up and turning left, the synthesized videos usually have  poor visual quality. However, even if a synthesized video is unreal to humans, it can still bypass the current anti-deepfake detection mechanism with a very high success rate. For example, the result of FaceShifter in Figure \ref{fig:action_evaluation} shows that the anti-deepfake evasion rate on TC can still achieve as high as 100\%. Therefore, the current anti-deepfake detection should be further improved to enhance the security of the target API.




	

\paragraphbe{Security of Different Actions and Lengths}
An action-based FLV API usually supports different actions and action sequence lengths. Then, we evaluate the security variance for different action requirements, and find that they do not result in security variance. Due to the space limitations, more details are deferred to Appendix \ref{appendix_aflv}.

\section{Exploration}
\label{exploration}
In this section, we explore various factors that may affect the attack effectiveness from the perspective of deepfake in-depth and discuss potential improvements in bypassing FLV\footnote{Withou explicit specification, the exploration experiment is conducted on silence-based FLV.}. To better demonstrate the impacts of various factors, we consider the two most effective methods (FaceShifter and FOMM) during the exploration. The overview of insights found in this section is shown in Remarks 5 - 8 of Figure \ref{figure:overview}, which correspond to the core insights found in the exploration on the bias of API, adversarial training, influence factors, and the customized two-stage attack, respectively.

\begin{table}[t]{\footnotesize
\setlength{\tabcolsep}{1pt}

	\centering

\begin{tabular}{cccccccc}
Platform                 & \multicolumn{2}{c}{Attributes}    & \begin{tabular}[c]{@{}c@{}}Liveness\\ Evasion \end{tabular} & $P$-value                    & \begin{tabular}[c]{@{}c@{}}Anti-deepfake\\ Evasion \end{tabular} & $P$-value                 & \begin{tabular}[c]{@{}c@{}}Overall\\ Evasion \end{tabular} \\ \hline
\multirow{4}{*}{BD}   & \multirow{2}{*}{Race}   & Colored & 82\%                                                           & \multirow{2}{*}{\textbf{0.049}}     & \textbf{72\% }                                                               & \multirow{2}{*}{\textbf{0.0055}} & 63\%                                                          \\
                         &                         & White   & \textbf{89\%}                                                          &                            & 58\%                                                                &                         & 55\%                                                          \\
                         & \multirow{2}{*}{Gender} & Male    & 74\%                                                           & \multirow{2}{*}{\textbf{0.0001}}    & 59\%                                                                & \multirow{2}{*}{0.53}   & 50\%                                                          \\
                         &                         & Female  & \textbf{96\%}                                                           &                            & \textbf{62\%}                                                                &                         & 62\%                                                          \\ \hline
\multirow{4}{*}{TC} & \multirow{2}{*}{Race}   & Colored & 43\%                                                           & \multirow{2}{*}{\textbf{0.00058}}   & \textbf{93\% }                                                               & \multirow{2}{*}{0.78}   & 42\%                                                          \\
                         &                         & White   & \textbf{67\% }                                                          &                            & 92\%                                                                &                         & 63\%                                                          \\
                         & \multirow{2}{*}{Gender} & Male    & 46\%                                                           & \multirow{2}{*}{\textbf{0.0000016}} & 96\%                                                                & \multirow{2}{*}{0.31}   & 41\%                                                          \\
                         &                         & Female  & \textbf{78\%}                                                           &                            & \textbf{99\% }                                                               &                         & 78\%                                                          \\ 
\end{tabular}

\caption{\footnotesize Evaluation of bias and statistical tests.}
\label{bias}}	
	
\end{table}

\subsection{Bias of API}
To explore the bias in FLV, we first divide the video dataset into two groups according to a particular attribute that might have bias. Then, we  sample 100 videos from each group and use \texttt{LiveBugger} to directly evaluate a target API with these sampled videos (not use  \textit{Deepfaker Engine}). As the video dataset provides a large number of live videos sampled from different environments (e.g., lighting), to limit the influence of other factors, we manually select two groups of videos to ensure that they have the same number of videos from the same environment. Considering the simplicity of distinguishing attributes, we mainly explore the potential bias in gender and race. In the future, we will explore the bias of more attributes in FLV.  We show the evaluation results and the corresponding $t$-test statistics in Table \ref{bias}, respectively. Note that,  since the overall evasion rate depends on liveness evasion rate and anti-deepfake evasion rate, we omit the statistical test for overall evasion rate.

From Table \ref{bias}, we can see the bias in FLV. 1)
The racial and gender biases exist in the presentation attack detection (measured by liveness evasion rate) of FLV API, and all the $P$-values for liveness evasion rate are less than 0.05 (many of them are even less than 0.01), which means that such biases are significant. For example, the liveness evasion rate of males on TC is only 46\%, while that of females can achieve as high as 78\%, and the corresponding $P$-value is 0.0000016, which indicates a significant gender bias. 
2) Although the bias of anti-deepfake evasion rate is not significant as that of the liveness evasion rate, it also exists in some cases. For example,  the anti-deepfake evasion rate of the colored on BD can achieve as high as 72\%, while that of the white is only 58\%, and the $P$-value is 0.0055, which means a statistically significant bias.  


In summary, there are biases in FLV, which may bring significant security risks to a particular group of people. How to eliminate the biases in FLV is an interesting future work.



\subsection{Adversarial Training and Anti-deepfake Detection}
\label{adv}

According to Figures \ref{fig:silent_evaluation}, \ref{fig:voice_evaluation} and \ref{fig:action_evaluation}, compared to FOMM, FaceShifter can bypass the anti-deepfake detection more efficiently. We speculate that this is due to the adversarial training used in FaceShifter. Specifically, in adversarial training (widely used in Generative Adversarial Networks \cite{goodfellow2014generative}),  the goal of the discriminator is to distinguish the synthesized videos from real ones, which is similar to the goal of anti-deepfake detection, while the goal of the generator is to deceive the discriminator. Therefore, the adversarial training may make the synthesized video more likely to bypass the anti-deepfake detection (i.e., higher anti-deepfake evasion rate). Note that, without the discriminator, FaceShifter fails to synthesize satisfying videos. We thus utilize FOMM to explore the role adversarial training (i.e., discriminator) for bypassing FLV. We present the results and the corresponding $t$-test statistics in Table \ref{adversarial_training}. 


\begin{table}[t]
	{\footnotesize
\setlength{\tabcolsep}{1pt}
	\centering

\begin{tabular}{ccccccc}
Platform                 & Method     & \begin{tabular}[c]{@{}c@{}}Liveness\\ Evasion\end{tabular} & $P$-value               & \begin{tabular}[c]{@{}c@{}}Anti-deepfake\\ Evasion\end{tabular} & $P$-value                     &  \begin{tabular}[c]{@{}c@{}}Overall\\ Evasion\end{tabular} \\ \hline
\multirow{2}{*}{BD}   & FOMM       & 95\%                                                           & \multirow{2}{*}{0.14} & 33\%                                                                & \multirow{2}{*}{\textbf{0.00000039}}                                                      & 33\%                                                          \\
                         & FOMM (Adv) & \textbf{97\%}                                                           &                       & \textbf{46\%}                                                                &                             &  \textbf{46\%}                                                          \\
\multirow{2}{*}{TC} & FOMM       & 29\%                                                           & \multirow{2}{*}{0.29} & 59\%                                                                & \multirow{2}{*}{\textbf{0.042}}      &  19\%                                                          \\
                         & FOMM (Adv) & \textbf{36\%}                                                           &                       & \textbf{69\% }                                                               &                             &  \textbf{25\% }                                                       \\ 
\end{tabular}
\caption{\footnotesize Evaluation of adversarial training.}
	\label{adversarial_training}}
\end{table}

According to Table \ref{adversarial_training}, adversarial training significantly improves the attack effectiveness of FOMM, especially in terms of bypassing the anti-deepfake detection. For example, after adversarial training, the anti-deepfake evasion rate is increased from 33\% to 46\% on BD, and the corresponding $P$-value is 0.00000039, which indicates a significant improvement. In addition, the adversarially trained FOMM is also more effective in terms of other metrics, although the improvement is not as significant. In summary, even though no visible visual difference exists between the videos synthesized by normal and adversarially trained FOMM, the latter can bypass FLV more effectively.


\subsection{Driving Videos}
\label{exp_driving}

 Intuitively, whether a driving video passes FLV or not may affect the success of the synthesized video. To validate this, we utilize \texttt{LiveBugger} to evaluate the role of the driving video for bypassing FLV. Specifically, we first sample 20 images from the image dataset as the target images. Then, we configure the driving videos in \texttt{LiveBugger} as videos that can pass silence-based FLV and cannot pass it, respectively, and utilize \texttt{LiveBugger} to 
evaluate a target FLV API.


\begin{table}[t]{\footnotesize
\setlength{\tabcolsep}{1pt}

	\centering


\begin{tabular}{cccccc}
\multirow{2}{*}{Platform} & \multirow{2}{*}{Method} & \multicolumn{4}{c}{Successful / Failed Image}                                                                                                                                                                                                                         \\ \cline{3-6} 
                          &                         & \begin{tabular}[c]{@{}c@{}}Liveness \\ Evasion \end{tabular} & \begin{tabular}[c]{@{}c@{}}Anti-deepfake \\ Evasion \end{tabular} & \begin{tabular}[c]{@{}c@{}}Face\\ Matching \end{tabular} & \begin{tabular}[c]{@{}c@{}}Overall\\ Evasion \end{tabular} \\ \hline
\multirow{2}{*}{BD}    & FOMM                    & 100\% / 87\%                                                       & 49\% / 29\%                                                            & 100\% / 100\%                                               & \textbf{49\% / 26\%}                                                     \\
                          & FaceShifter             & 96\% / 95\%                                                        & 57\% / 65\%                                             & 94\% / 95\%                                                 & 54\% / 60\%                                                     \\
\multirow{2}{*}{TC}  & FOMM                    & 32\% / 5\%                                                         & 55\% / 58\%                                                            & 99\% / 100\%                                                &\textbf{27\% / 5\% }                                                     \\
                          & FaceShifter             & 60\% / 50\%                                                        & 100\% / 100\%                                                          & 93\% / 96\%                                                 & 58\% / 50\%                                                     \\ 
\end{tabular}
	\caption{\footnotesize Evaluation of the target image.\label{target_images}}}
\end{table}

\begin{figure}[t]
	\centerline{\includegraphics[width=0.9\columnwidth]{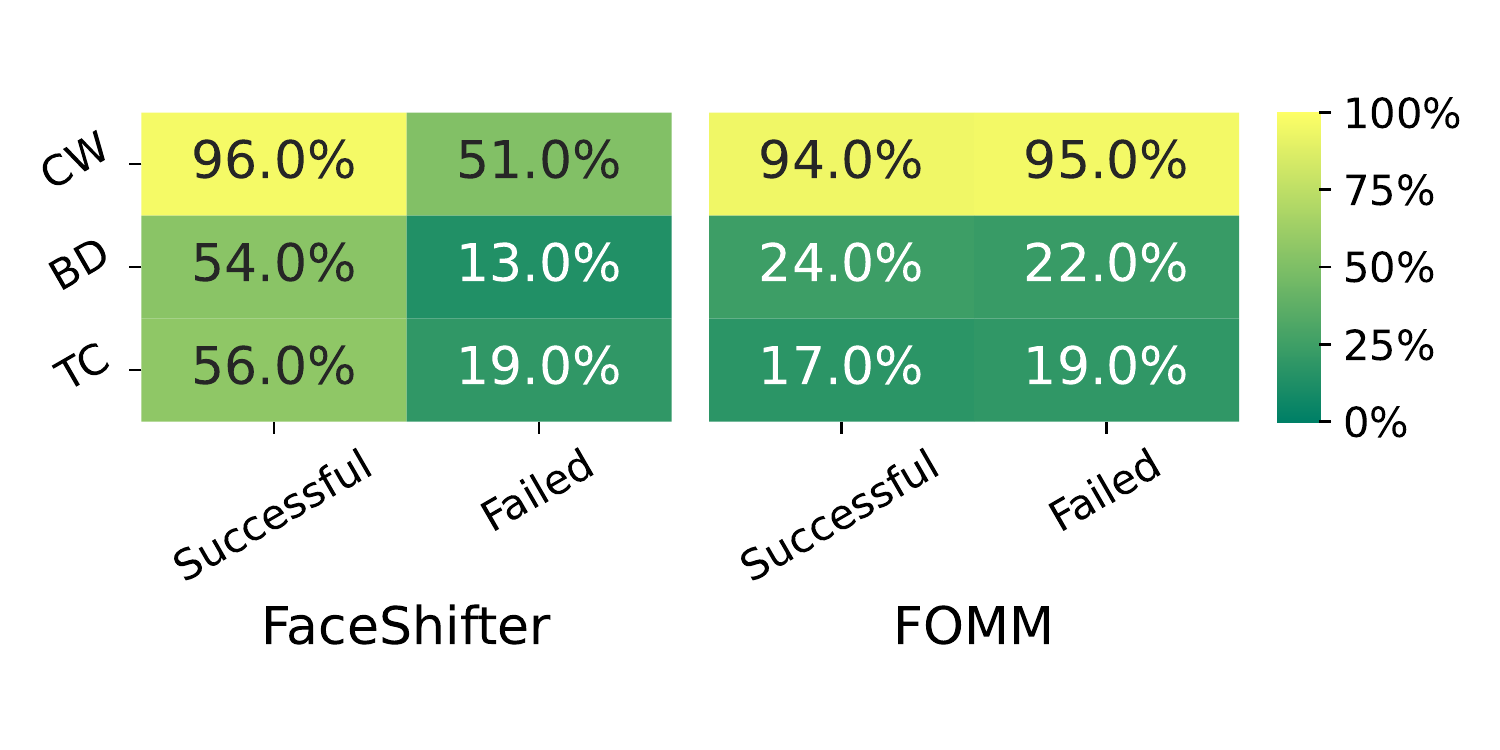}}
	\caption{\footnotesize Evaluation  of the driving video.  }	
	\label{figure:exp_driving}
\end{figure}

Figure \ref{figure:exp_driving} shows the influence of the driving video on the bypass rate, where the driving video that can pass silence-based FLV is denoted as `successful' and `failed' otherwise. From Figure \ref{figure:exp_driving}, we can see that the influence of the driving video on FaceShifter is more significant than that on FOMM. For example, FaceShifter with the passed driving video can achieve an overall evasion rate of 96\% on CW, while with the failed driving video, it can only achieve an overall evasion rate of 51\%. As for FOMM, the successful and failed driving videos result in similar overall evasion rate. After analysis, we speculate that this is because FaceShifter is a face-swapping method, while FOMM is a face reenactment method. The face-swapping method  swaps the identity (face) of the driving video with the identity (face) of the target image. The obtained video is more similar to the driving video, including background, expression, and posture. Therefore, the driving video has a significant influence on face swapping methods. On the contrary, face reenactment methods fuse the action of the driving video to the target image to reenact the target image, i.e., the synthesized video is more similar to the target image. Therefore,  face reenactment methods are less affected by the driving video.


\subsection{Target Images}

Based on the analysis in Section \ref{exp_driving}, intuitively, the target image may greatly influence face reenactment methods for bypassing FLV  compared to  face swapping methods. Now, we discuss the influence of the target image on the bypass rate. For exploring its influence, we randomly sample 20 images that can pass image-based FLV from the image dataset and 20 images that cannot pass, respectively. Then, we use \texttt{LiveBugger} to explore such influence. The evaluation results are shown in Table \ref{target_images}.

As shown in Table \ref{target_images},  the target image has different influences on different deepfake techniques for bypassing FLV. Specifically, for FOMM, the synthesized video based on the passed target image  is more likely to bypass FLV than the failed one. For instance, for FOMM, the successful image can achieve an overall evasion rate  of 49\% on BD, much higher than that (26\%) achieved by the failed image. While for FaceShifetr, the successful and failed images achieve similar overall evasion rate. Note that, FaceShifter and FOMM are face-swapping and face-reenactment methods, respectively. We conjecture that this is because each frame in a synthesized video by a face reenactment method is more similar to the target image, such as background and identity. Therefore, the target image has more impact on face reenactment methods.

\begin{table}[t]{\footnotesize
\setlength{\tabcolsep}{1pt}
	\centering

\begin{tabular}{ccccc}
Platform                 & Method                   & \begin{tabular}[c]{@{}c@{}}Liveness \\ Evasion \end{tabular} & \begin{tabular}[c]{@{}c@{}}Anti-deepfake\\ Evasion \end{tabular} &  \begin{tabular}[c]{@{}c@{}}Overall\\ Evasion \end{tabular} \\ \hline
   & FOMM                     & 91\% / 10\%                                                         & 22\% / 52\%                                                                                                                & \textbf{22\% / 5\%}                                                         \\
   BD                       & FOMM (Stage1)           & 100\% / 14\%                                                          & 86\%  / 94\%                                                                                                            & \textbf{86\% / 13\%}                                                        \\
    / TC                     & FOMM (Stage2)           & 94\%  / 10\%                                                         & 39\%  / 52\%                                                                                                                 & \textbf{39\% / 6\%}                                                          \\
                         & \textbf{FOMM (Stage1+Stage2)} & 100\% / 20\%                                                          & 92\% / 99\%                                                                                                                   & \textbf{92\% / 20\%}                                                          \\ 
\end{tabular}
\caption{\footnotesize Evaluation of the two-stage approach.}
\label{two_stage}}
\end{table}

\subsection{A Customized Two-stage Attack}
\label{improve_attack}

According to the above interesting insights, the driving video and the target image can improve the attack effectiveness of face swapping methods and face reenactment methods, respectively. For face swapping, since the driving video is under full control of an adversary, he/she can adopt the passed driving video to improve the attack effectiveness; while for face reenactment, the obtained target image is not under full control of an adversary. To this end, we propose a two-stage approach to improve the performance of face reenactment methods further.

\textbf{Stage 1: } After we get a target image, we transform it to an image that can pass image-based FLV, which helps synthesize a video that can bypass silence-based FLV.  For simplicity, we use FaceShifter to transform the image, which can be trivially extended to other face swapping methods.

\textbf{Stage 2: }
According to the analysis in Section \ref{adv}, adversarial training can improve the effectiveness of bypassing FLV. Therefore, at this stage, we use a face reenactment method that incorporates adversarial training to synthesize fake videos.


Now, we explore the effectiveness of the customized two-stage attack. Specifically, we first  randomly select 20 target images and transform them with FaceShifter to make them pass  image-based FLV. Then, due to the effectiveness of FOMM for bypassing FLV, we utilize it to synthesize videos based on five driving videos and the transformed images. Finally, we use the synthesized videos to evaluate the target FLV API. The evaluation and ablation studies of the customized two-stage approach are shown in Table \ref{two_stage}.

From Table \ref{two_stage},  we have the following observations. 1) Both Stage 1 and Stage 2 can improve the effectiveness of bypassing FLV. For example, the original FOMM can achieve an overall evasion rate of 22\% on BD, while the FOMM with Stage 1 and Stage 2 can improve the overall evasion rate to 86\% and 39\% respectively, which indicates a big improvement. 2) The two-stage approach (Stage 1 + Stage 2) achieves the highest overall evasion rate. For example, it can increase the overall evasion rate on BD to 92\%. These observations further confirm the insights observed in previous sections.

\begin{table}[t]{\footnotesize
\setlength{\tabcolsep}{1pt}
	\centering
	
\begin{tabular}{cccccc}
Name          & Type  &  FLV  & Area              &  Result &  \\ \hline
 BI (Basic)       & \multirow{2}{*}{API}  & Silence-based & \multirow{2}{*}{Germany}      & 84\%       \\
 BI (with Assistance)  &  & Action-based  &       & 82\%        \\
 PI   & API   & Image-based  &   French         & 90\%      \\
 AC   & API  & Image-based  & Korea & 52\% \\
NT       & Demo APP & Silence-based&  Lithuania   & \checkmark       \\
  BD   & Demo APP & Action-based  & China  &  \checkmark        \\
   TC   & Demo APP& Silence-based  &  China  & \checkmark         \\
  FPP   &   Demo  APP& Action-based  &   China    & $\times$        \\ 
\end{tabular}

\caption{\footnotesize Evaluation of FLV APIs from the global market. $\checkmark$ denotes a successful attack; $\times$ denotes a failed attack.}
\label{global_result}}
\end{table}

\section{Evaluation on Global FLV Services}
\label{global}
In Section \ref{evaluation}, we have evaluated the most representative FLV APIs in China. Different from the AI cloud vendors evaluated in Section \ref{evaluation}, most of the global FLV vendors often provide a specific type of FLV.  For better representing the threat surface, we also utilize \texttt{LiveBugger} to evaluate these leading global vendors.
According to the FLV service forms, we evaluate them in the following ways. For the vendors that provide the FLV API, we directly evaluate them in a way similar to Section \ref{evaluation}. For some vendors, we only have access to their demo APPs, which limits the flexibility for calling the low-level FLV API. Thus, we evaluate them in a more real-world setting. Specifically, we first hijack the camera video stream of a local device, which runs the evaluated demo APP.
Since evaluating the demo APPs in a large-scale manner is challenging, in the evaluation, we randomly select five identities from the image dataset as the victims. Then, we utilize \texttt{LiveBugger} to synthesize the corresponding fake videos and push them into the demo APPs in a real-time manner. For each demo APP, we consider the attack successful if more than three out of five trials are successful.

We evaluate the APIs or demo APPs provided by representative vendors from the global market. The reason to consider these vendors is as follows:  1) BI is one of the leading biometric vendors, which was successfully tested for level A and level B attacks according to  ISO 30107-3; 2) PI was selected as a finalist for the ``Best Use of AI in fintech" in IFTA2020; 3) NT is among the eight most accurate face recognition algorithm vendors; 4) AC is one of the leading visual recognition AI firms in Korea; 5) FPP is widely evaluated by previous work\cite{sharif2016accessorize}; 6) due to the large market share of BD and TC, we further evaluate their demo APPs' vulnerability. We present the service types provided by these vendors and the evaluation results in Table \ref{global_result}. Note that, we report attack success rate and  binary attack result (i.e., a successful or failed attack) for APIs and demo APPs, respectively.


From Table \ref{global_result}, we can see that the security risks also exist in FLV vendors from the global market. (1) All the evaluated APIs can be bypassed effectively. For example, as shown in Table \ref{global_result}, we can achieve an attack success rate of 90\% on PI. (2) Besides, compared to the basic FLV provided by BI (i.e., silence-based FLV), the FLV with assistance (i.e., action-based FLV) brings limited security gain, which is the same as the observations in Section \ref{evaluation}. (3) For demo APPs, most of them can be attacked successfully, further threatening the security of downstream clients. As for FPP, since it deploys coherence detection, the synthesized video via the stitched driving video can be detected with high confidence. Therefore, we alternatively manually record the required driving video. Although we cannot fully bypass it in this way, we decrease the confidence that a fake video is detected as an attack from 0.99 to around 0.5 (the detection threshold).

In summary, similar to the evaluation results in Section \ref{evaluation}, the evaluation on the global FLV vendors confirms that most of them are vulnerable to deepfake-powered attacks, which is a severe and widely existing threat.

\section{Proof-of-concept Attack}
\label{proof_of_concept}
As the source of the software supply chain, the security risks of APIs will threaten many downstream applications and clients. In previous sections, we have illustrated that the security risks are widespread in various FLV APIs and demo APPs. In this section, we evaluate such risks in a more real-world setting.
We conduct proof-of-concept attacks on representative clients of these APIs via a manner similar to the demo APP evaluation in Section \ref{global}, demonstrating the feasibility of conducting such attacks in the real world. Specifically, we hijack the camera video stream of the evaluated applications and feed them with the synthesized video stream in a real-time manner.
The evaluated applications are selected from the representative clients of the corresponding FLV vendors according to their official websites, including HN Airlines, TK Insurance, R360 and HZ Citizen Card. HN Airlines is rated as Skytrax five-star airline and one of the best airlines in China.  R360 is one of the most valued fintech unicorns in the world. TK Insurance is one of the largest life insurers by premium income in China. HZ Citizen Card is a widely-used government service application in one of the smartest cities in China. All of these applications have a vast amount of users.  For each APP, we recruit five volunteers from the university as its users. Note that one volunteer may use multiple APPs. These volunteers register accounts for the corresponding APPs, which can be considered as an enrollment process. All the APPs provide services that require FLV. Then, we use the volunteers’ accounts (authorized by them) to evaluate the security of FLV services. We present the evaluation results in Table \ref{poc_result}. 


From  Table \ref{poc_result}, we can see that  all the evaluated APPs can be attacked successfully, and thus threaten the security of  millions of  users of these APPs. Taking HN Airlines (a representative client of BD's FLV services) as an example, we show the attack screenshot in Figure \ref{fig:hainan} of Appendix \ref{appendix_other}.
As shown in Figure \ref{screenshot_shakehead}, the application requires the user to shake his head. Therefore, we reenact the target facial image to shake his head, which can be recognized by the application. We repeat the above reenactment process for each of the required actions, and then the corresponding verification can be bypassed. It makes compromising the account of the target user possible, e.g.,  stealing his/her accumulated miles. 

\begin{table}[t]{\footnotesize
\setlength{\tabcolsep}{2pt}
	\centering
	
\begin{tabular}{cccc}
Type & Name      & Attack Result & Users/Downloads \\ \hline
\multirow{4}{*}{APPs} & HN Airlines   & \checkmark    & 30 million      \\
                     &  R360        & \checkmark    & 0.33 billion    \\
                     &  TK Insurance  & \checkmark    & 15 million      \\
                     &  HZ Citizen Card  & \checkmark    & 35 million      \\
\end{tabular}
\caption{\footnotesize Evaluation of proof-of-concept attacks. \checkmark denotes a successful attack; $\times$ denotes a failed attack.}
\label{poc_result}}
\end{table}



\section{Discussion}
\label{discussion}

\subsection{Ethical Consideration}
In this work, we conduct a comprehensive security evaluation on FLV using deepfake, which may raise some ethical concerns. Similar to the previous studies about the security of AI-powered systems \cite{zhao2020large, shi2021adversarial, zhao2018towards}, we pay special attention to the legal and ethical boundaries. First, we use open-source datasets to conduct deepfake synthesis and security evaluation, which is a legitimate and common practice in face-related security research \cite{shan2020fawkes,li2019faceshifter}. Besides, since we directly call the target APIs with the target image and the synthesized images/videos, no fake accounts were created for the celebrities.
Second,  our evaluation of the commercial FLV APIs strictly follows the official instructions, and we paid for the API usage. Besides, we limit the Queries Per Second (QPS) to the recommended value. Therefore, our evaluation does not affect the normal service of the target vendor. For demo APPs provided by some vendors, since it is provided for trial use, the evaluation on them will not affect the normal business affairs. Thirdly,  for the proof-of-concept attacks, we evaluate some widely used applications with the accounts of volunteers and get their authorization, which does not affect other users and the business affairs of the corresponding company.  Besides, we have reported our results to the corresponding vendors and got their acknowledgments. Finally, we replace the vendor and APP name with cryptonyms, which can minimize the potential security risks to the affected vendors.

\subsection{Vulnerability of FLV Services}
In this work, we utilize deepfake-powered attacks to evaluate the vulnerability of existing representative FLV APIs and find that almost all of them can be compromised. As the source of the FLV service supply chain, the vulnerability will be inherited by downstream APPs, further threatening millions of end-users. After fine-grained analysis, we speculate that the following reasons cause it. (1) The design of the verification process of some FLV APIs is problematic. For example, the voice-based FLV API of BD detects the audio signal and facial liveness in a separate manner, which is highly vulnerable. We can easily bypass it via importing the required audio to the video. (2) The effectiveness of FLV's defense mechanism is concerning. Specifically, it seldom considers stronger attacks (e.g., the proposed deepfake-powered attacks). For comparison, we evaluate the effectiveness of FLV APIs for defending against the presentation attacks and show the results in Table \ref{presentation_attack} of Appendix  \ref{appendix_other}. It can be seen that  the presentation attack can hardly pose any threat to the current FLV APIs. However, as we can see from Section \ref{evaluation}, the effectiveness of the deepfake-powered attack is much higher than that of  the presentation attack. Therefore, it is urgent for FLV designers to integrate the defense capability of FLV against stronger attacks, especially the new arising attacks.

In summary, the imperfection of underlying FLV mechanisms and the inadequate defense capabilities make current FLV services vulnerable to deepfake-powered attacks.

\subsection{Variations of Attack Effectiveness}
The evaluation in Section \ref{evaluation} shows that some deepfake methods show higher effectiveness to evade FLV. For instance, Figures \ref{fig:silent_evaluation} and \ref{fig:voice_evaluation} show that FaceShifter and FOMM often attain higher overall evasion rate on different vendors compared with other deepfake methods. In general, more advanced deepfake methods (e.g., FaceShifter and FOMM) often obtain better visual results, leading to higher attack effectiveness.
Meanwhile, different deepfake methods also show variations across different vendors. Figure \ref{figure:radar_compare} indicates that FaceShifter performs better on BD and TC, while FOMM performs better on ST and iFT. Without access to the technical details of the target FLV vendors, we speculate that such variations are attributed to the defense measures deployed by different vendors.  For instance, certain vendors may deploy defenses against specific deepfake attacks.

\subsection{Security  Suggestions}
Below, we provide security  suggestions based on the valuable insights observed in Sections \ref{evaluation} and \ref{exploration}. Specifically, we provide customized suggestions for different types of FLV.
a) \textbf{Image-based FLV.} According to the evaluation in Section \ref{eval_iflv}, an adversary can utilize one facial image to bypass the FLV systems. We recommend abandoning image-based FLV in the future. b) \textbf{Silence-based FLV.} Since silence-based FLV does not require any auxiliary information, an adversary can easily utilize deepfake to bypass it. Therefore, anti-deepfake detection becomes necessary for silence-based FLV. However, a big gap exists between current anti-deepfake detection and human perception. Therefore, anti-deepfake detection should draw more research attention in the future. Note that,  anti-deepfake detection is also necessary for voice-based FLV and action-based FLV. c) \textbf{Voice-based FLV.}  Voice-based FLV can adopt a cross-modal manner in the future. Specifically, during verification, it can consider the match of lip movements with the audio signal, or even voiceprint to improve the security.  Besides, the form of the random process should not be limited to digits, but a random process with much more diversity.
d) \textbf{Action-based FLV.} Head movements often cause visual incoherence and unnatural distortion in the synthesized video. Therefore, coherence and anti-deepfake detection should play a vital role in developing secure action-based FLV in the future.  Besides, action-based FLV may adopt actions that are hard to be synthesized by deepfake. 




\subsection{Limitations and Future Work}


Firstly, our goal is to evaluate the security of FLV. Therefore, we integrate several SOTA deepfake methods, which can efficiently bypass FLV.  Since many deepfake methods are based on a similar methodology, \texttt{LiveBugger} does not include all the deepfake methods.  However, thanks to its high extendability, \texttt{LiveBugger} is ready to be extended to incorporate new deepfake methods. Moreover, we plan to open-source \texttt{LiveBugger} to facilitate the FLV security research and encourage the community to contribute more techniques. 



Secondly, in this study, we mainly focus on the one-shot setting. Indeed, an adversary usually can obtain more than one facial image of the victim, which may bring more security risks to FLV. Note that, few-shot deepfake methods \cite{wang2019few,zakharov2019few} can be easily incorporated into \texttt{LiveBugger} if the threat model is relaxed to the few-shot setting. According to \cite{wang2019few,zakharov2019few}, compared to the one-shot setting, the few-shot setting can output more realistic results, which may pose greater threats to FLV. We plan to extend our work to the few-shot setting in the future.

Finally, extending the current work to other domains, such as speaker recognition, is an interesting future work. Besides, according to the security suggestions, developing effective and robust defense schemes is also a promising future work.

\section{More Related Work}


 In Black Hat 2009, researchers first showed how to easily bypass facial authentication using one facial image \cite{Duc2009YourFI}.  Later, based on the facial disclosure shared on social networks,  Li \textit{et al.}  systematically studied the threat brought by the  presentation attack \cite{li2014understanding}. 
 
 To mitigate such attacks, many defenses have been proposed via  \cite{tan2010face,maatta2011face,boulkenafet2015face,benlamoudi2015face,parveen2016face}. In early times,  researchers used hand-crafted features to detect face spoofing. For example, eye blinking detection is a common heuristic used by many FLV systems \cite{pan2007eyeblink}.  Later on, with the rapid progress of deep learning, many researchers used deep features to detect the presentation attack. Jorabloo \textit{et al.}  proposed a CNN  architecture with proper constraints and supervisions for decomposition to detect fake faces \cite{jourabloo2018face}. George \textit{et al.} also utilized CNN for face spoofing detection with deep pixel-wise supervision \cite{george2019deep}. Recently, Spatio-Temporal Anti-Spoof Network (STASN) achieved SOTA  performance on public anti-spoofing datasets  \cite{yang2019face}. Except for these detection methods, researchers also proposed many defenses from the perspective of FLV design \cite{chetty2006multi, choudhury1999multimodal, frischholz2003avoiding}.  Chetty \textit{et al.} proposed a challenge-response-based liveness detection mechanism that involves user interaction (speaking given digits), which can significantly improve FLV security \cite{chetty2006multi}.  More recently, Tang et al. proposed a liveness detection protocol based on light reflections \cite{TangZZZ18}. It requires the screen emitting light of random colors and uses a camera to capture the light reflected from the face as the liveness clue. Uzun et al. presented a Captcha-based liveness detection system, which requires the user to record a video when answering a Captcha to complete the verification \cite{UzunCEL18}. 
 
 
Compared to existing studies, our work differs in several major aspects. 1) Most previous work focuses on developing new liveness detection mechanisms \cite{chetty2006multi, TangZZZ18, UzunCEL18}; in contrast, our work aims to raise concerns about the change of attack surfaces caused by deepfake and shed light on the future directions of improving the security assurance of current FLV services.
2) Prior work evaluates face recognition without liveness detection. For example, the services evaluated in Uzun’s work \cite{UzunCEL18}, including Face API MS Azure and Amazon Rekognition, do not assume liveness detection capabilities. Although Uzun et al. used smile detection as a liveness clue, it is not officially provided by the above services. Thus, it dose not fully expose the security risks of the latest FLV services enhanced by liveness detection. 3) The attack-defense landscape of FLV has since changed significantly. On one hand, FLV vendors have greatly improved their security. For instance, some vendors claim that their services are equipped with deepfake detection capabilities. On the other hand, recent years have witnessed striding advances in deepfake techniques, which pose unprecedented challenges for FLV. Therefore, it is imperative to re-evaluate the security assurance of the latest FLV services facing SOTA deepfake techniques. In this paper, to bridge the gap, we conduct the first systematic evaluation and exploration of the threats of deepfake against FLV.

\section{Conclusion}
We design and implement \texttt{LiveBugger}, a first-of-its-kind security evaluation framework for FLV.  An extensive evaluation using \texttt{LiveBugger}
demonstrates that most representative FLV systems are highly vulnerable to deepfake-based attacks. Further, from the adversary's perspective, we explore the factors that may impact the attack effectiveness of deepfake. Based on the findings in this exploration, we propose a customized two-stage approach that can further boost the attack success rate by up to 70\%. To assess the threats in realistic settings, we perform proof-of-concept attacks in real-world applications. Lastly, we provide a set of suggestions to improve the security of FLV. We hope this work can shed light on developing more effective and robust FLV schemes.

\section*{Acknowledgements}
We thank our shepherd Bimal Viswanath and anonymous reviewers for their valuable feedback. This work was partly supported by the National Key Research and Development Program of China under No. 2020AAA0140004, National Natural Science Foundation of China (NSFC) under No. 62102360, U1936215 and U1836202, and the Zhejiang Provincial Natural Science Foundation for Distinguished Young Scholars under No. LR19F020003. Ting Wang was partially supported by the National Science Foundation under Grant No. 1951729, 1953813, and 1953893.

{\small \bibliographystyle{IEEEtran}
\bibliography{src}}

\appendix

\setlength{\textfloatsep}{2\baselineskip}
\setlength{\floatsep}{1\floatsep}
\setlength{\dblfloatsep}{1\dblfloatsep}
\setlength{\dbltextfloatsep}{1\dbltextfloatsep}
\setlength{\intextsep}{1\intextsep}
\setlength{\belowcaptionskip}{1pt}
\setlength{\abovecaptionskip}{1pt}

\section*{Appendix}

\section{Overall Setting}

  \subsection{Deepfake Methods}
  \label{appendix_deepfake}

 \texttt{LiveBugger} incorporates six SOTA deepfake methods, including X2Face \cite{wiles2018x2face}, ICface \cite{tripathy2020icface}, FSGAN \cite{nirkin2019fsgan}, First Order Method Model (FOMM) \cite{siarohin2019first} and FaceShifter \cite{li2019faceshifter}. We present their categories in Table \ref{deepfake}.  Below, we give them a brief introduction respectively.
 
 \textbf{X2Face} uses an embedding network  and a driving network to generate fake videos. The embedding network maps pixels from  the source frame/image to the embedded face, which can provide identity information. Then, based on the driving frame/image, the driving network maps pixels from the embedded face to the generated frame, which has the identity of the source frame/image and the pose/expression of the driving frame/image \cite{tripathy2020icface}.

 \textbf{ICface} is a two-stage generative adversarial network (GAN) based model trained in a self-supervised manner,  which can use human interpretable signals (e.g., head pose angles) to control the pose and expressions of a given face image \cite{tripathy2020icface}. 

\textbf{FSGAN} is a GAN-based approach that can be used for face swapping (FSGAN$_S$) and face reenactment (FSGAN$_R$). It first uses a recurrent reenactment generator to estimate the reenacted face and its segmentation from the source frame/image, and a segmentation generator to estimate the face and hair segmentation from the target frame/image. Then, based on the above information, it uses an inpainting generator to estimate the complete reenacted face. Finally, it uses a blending generator to completely blend the reenacted face and target face \cite{nirkin2019fsgan}.

\textbf{FOMM} first uses an unsupervised keypoint detector to extract first-order motion representation, including sparse keypoints and local affine transformations with respect to the reference frame/image. Then, the dense motion network uses the motion representation to generate dense optical flow from the driving frame/image to the source frame/image. Finally, the generator uses the source frame/image and the outputs of the dense motion network to generate the fake frame. Note that, the discriminator is optional during FOMM training \cite{siarohin2019first}.

\textbf{FaceShifter} is a novel two-stage GAN-based framework for high fidelity and occlusion aware face-swapping. It requires two input frames/images, i.e., a source frame/image to provide identity and a target frame/image to provide attributes (e.g., posture, scene lighting). In the first stage, it uses an Adaptive Embedding Integration Network (AEINet) to generate a high fidelity face-swapping result based on information integration (i.e., identity and attributes information). In the second stage, it
uses the Heuristic Error Acknowledging Network (HEARNet) to handle the facial occlusions and refine the result, and generate the final frame/image \cite{li2019faceshifter}. 

%

\begin{table}[h]
{\footnotesize
\setlength{\tabcolsep}{2pt}
	\centering
	
	\begin{tabular}{cc}
		Deepfake Method & Type \\ \hline
		X2Face \cite{wiles2018x2face}                  & Face Reenactment   \\ 
		ICface     \cite{tripathy2020icface}              &Face  Reenactment \\
		FSGAN$_S$   \cite{nirkin2019fsgan}              & Face Swapping      \\
		FSGAN$_R$   \cite{nirkin2019fsgan}             &Face  Reenactment   \\
		FOMM \cite{siarohin2019first}  &Face  Reenactment   \\
		FaceShifter   \cite{li2019faceshifter}           &Face  Swapping  \\ 
	\end{tabular}
	\caption{\footnotesize Deepfake methods used in  our work. }
	\label{deepfake}
	}
 \end{table}

  \section{Additional Evaluation}

\subsection{Silence-based FLV}
 \label{appendix_sflv}


\textbf{FaceShifter vs. FOMM.} Since FaceShifter and FOMM achieve a much higher overall evasion rate, we further compare them on different vendors. The evaluation results are shown in Figure \ref{figure:radar_compare}. It can be seen that FaceShifter and FOMM have different adaptability on different vendors.


\begin{figure}[h]
	\centering
	
	\begin{minipage}[t]{0.8\columnwidth}
		\centering
		\includegraphics[width=\columnwidth]{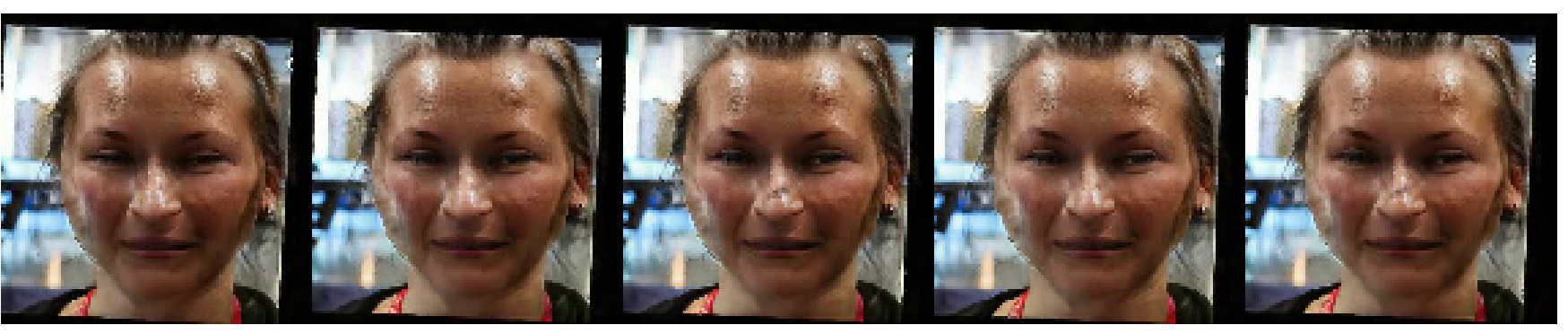}\\
		\label{image_faceswap}
	\end{minipage}
	
	\begin{minipage}[t]{0.8\columnwidth}
		\centering
		\includegraphics[width=\columnwidth]{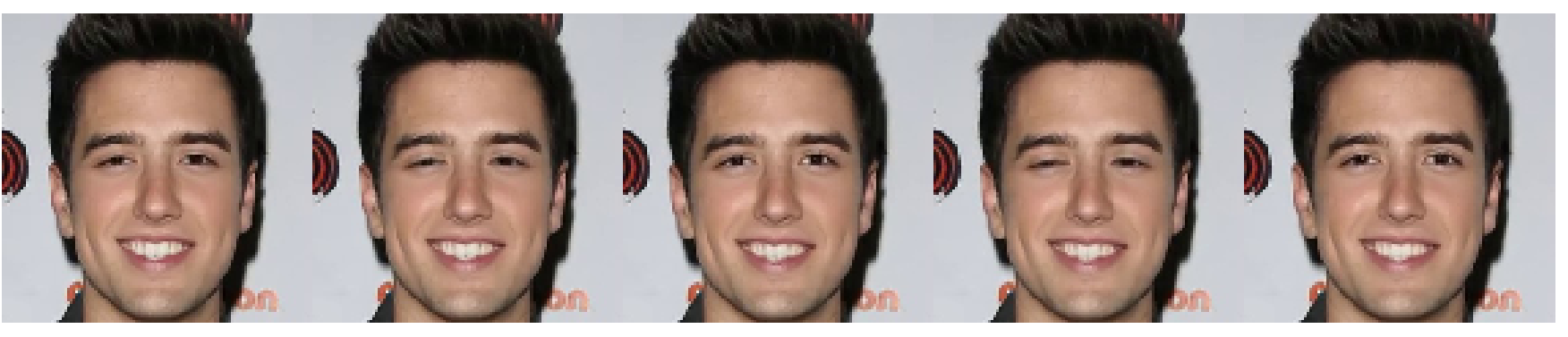}\\
		\label{image_facereen}	
	\end{minipage}%
	
	\centering
	\caption{\footnotesize The extracted frames from synthesized videos. The first row denotes frames extracted from videos that can pass deepfake detection, while the second row denotes frames extracted from videos that cannot pass deepfake detection.}
	\label{fig:extract_frame}
\end{figure}

\subsection{Voice-based FLV}
\label{appendix_vflv}



\begin{figure}[h]
	\centerline{\includegraphics[width=0.9\columnwidth]{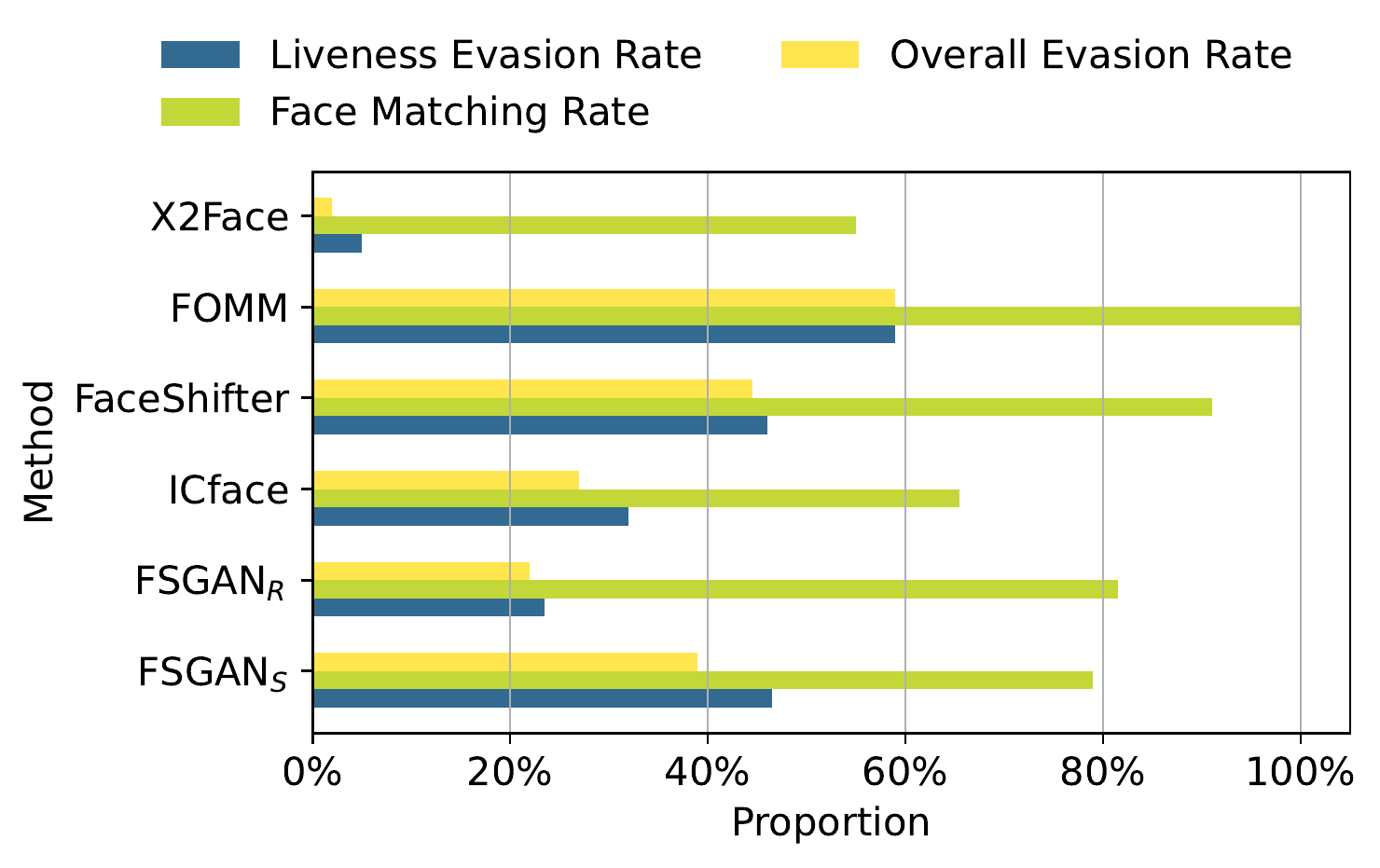}}
	\caption{\footnotesize Evaluation  with the customized driving video. Since CW has no anti-deepfake detection mechanism, we do not show its anti-deepfake evasion rate.}	
	\label{figure:voice_customized_video}
\end{figure}

\textbf{Lip Language Detection.} When the adversary obtains the given digits, he/she can interactively record a customized video with the matched lip movements as the driving video to synthesize the fake video. Below, we utilize the customized driving videos to evaluate CW's voice-based FLV API, as shown in Figure \ref{figure:voice_customized_video}.

From Figure  \ref{figure:voice_customized_video}, we can clearly see that even though the voice-based FLV API deploys lip language detection, it still suffers high risks. For example, FOMM can still achieve around 60\% overall evasion rate on CW. Therefore, although lip language detection brings security gain, it alone is not enough.

\begin{figure}[t]
	\centerline{\includegraphics[width=0.8\columnwidth]{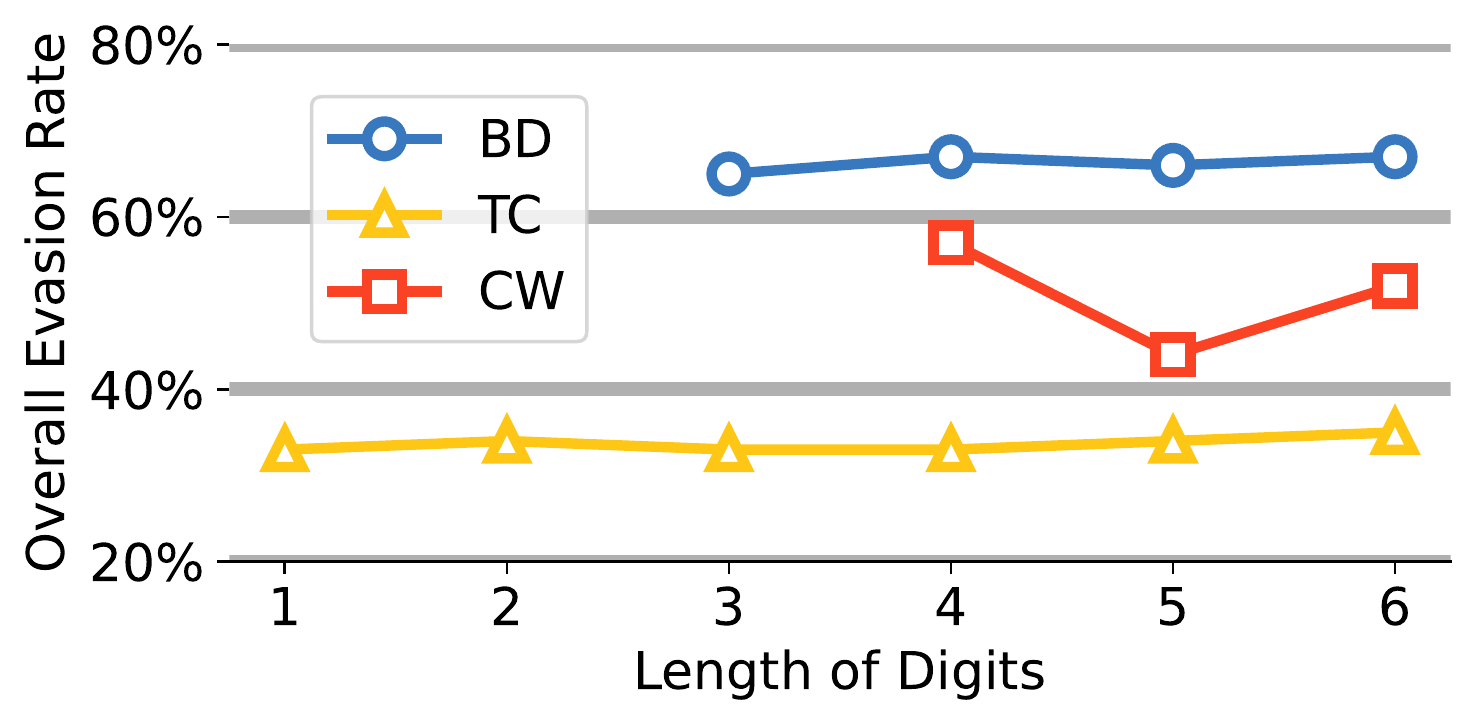}}
	\caption{\footnotesize Evaluation of the influence of the digit length on the voice-based FLV API.}	
	\label{figure:image_length}
\end{figure}

\textbf{Length of Given Digits.} We evaluate the influence of the digit length on the security of FLV on BD, TC, and CW, since they support changeable length. As CW deploys the lip language detection, we evaluate it with the customized driving videos. For the overall evasion rate under each length, we use the highest overall evasion rate that the deepfake methods can achieve. We present the evaluation results in Figure \ref{figure:image_length}.

From Figure \ref{figure:image_length}, we can clearly see that increasing the length of the digits at the cost of utility does not improve the security of a voice-based FLV API. For example, for BD and TC, the overall evasion rate barely changes. This is because the length of the digits does not influence voice recognition. While for CW, even though it deploys the lip language detection, since the driving video has matched lip movements, the influence of the digit length is limited if the deepfake method is proper in synthesizing the lip movements. The observation further illustrates that the current implementation of voice-based FLV may be problematic.

\subsection{Action-based FLV}
\label{appendix_aflv}

\begin{figure}[t]
	\centerline{\includegraphics[width=0.8\columnwidth]{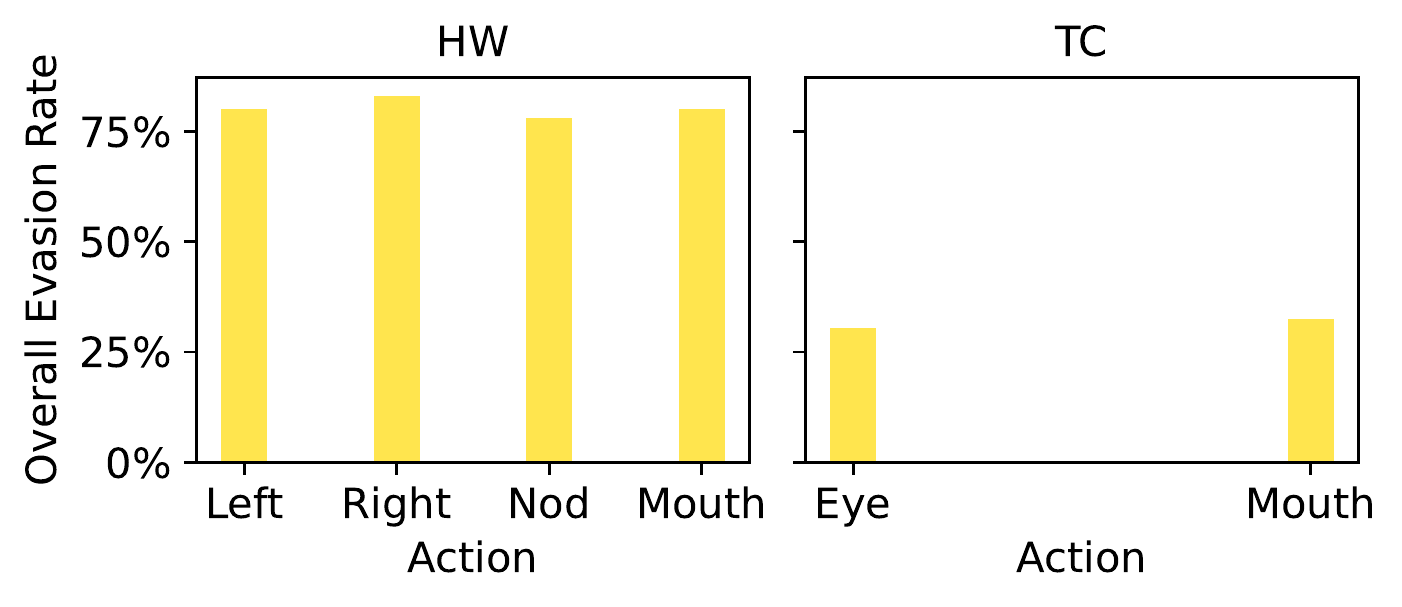}}
	\caption{\footnotesize Evaluation towards different actions.}
	\label{fig:action_type_evaluation}
\end{figure}

 \textbf{Security of Different Actions.} An action-based FLV API usually supports many actions, including blink, looking up, turning right, etc. In this way, the target API can randomly select a sequence of the supported actions. This random process is expected to improve the security of action-based FLV. Intuitively, different actions might have different security guarantees since different synthesis difficulties. To this end, we utilize \texttt{LiveBugger} to evaluate the security of different actions and show the evaluation results in Figure \ref{fig:action_type_evaluation}.


From Figure \ref{fig:action_type_evaluation}, we can find that the overall evasion rate of different actions for the same vendor is similar. For example, the result on HW in Figure \ref{fig:action_type_evaluation} shows that the overall evasion rate of all actions is around 80\%, which indicates that the security guarantees of different actions are not much different. 
Since the visual effect of the synthesized videos with actions involving large movements is much worse than that involving small movements, the former should be detected easily.
However, the target APIs do not behave differently. One reason could be that the target APIs do not deploy the coherence detection. Overall, the anti-deepfake detection ability in current action-based FLV APIs needs to be significantly improved.

\begin{figure}[t]
\centerline{\includegraphics[width=0.8\columnwidth]{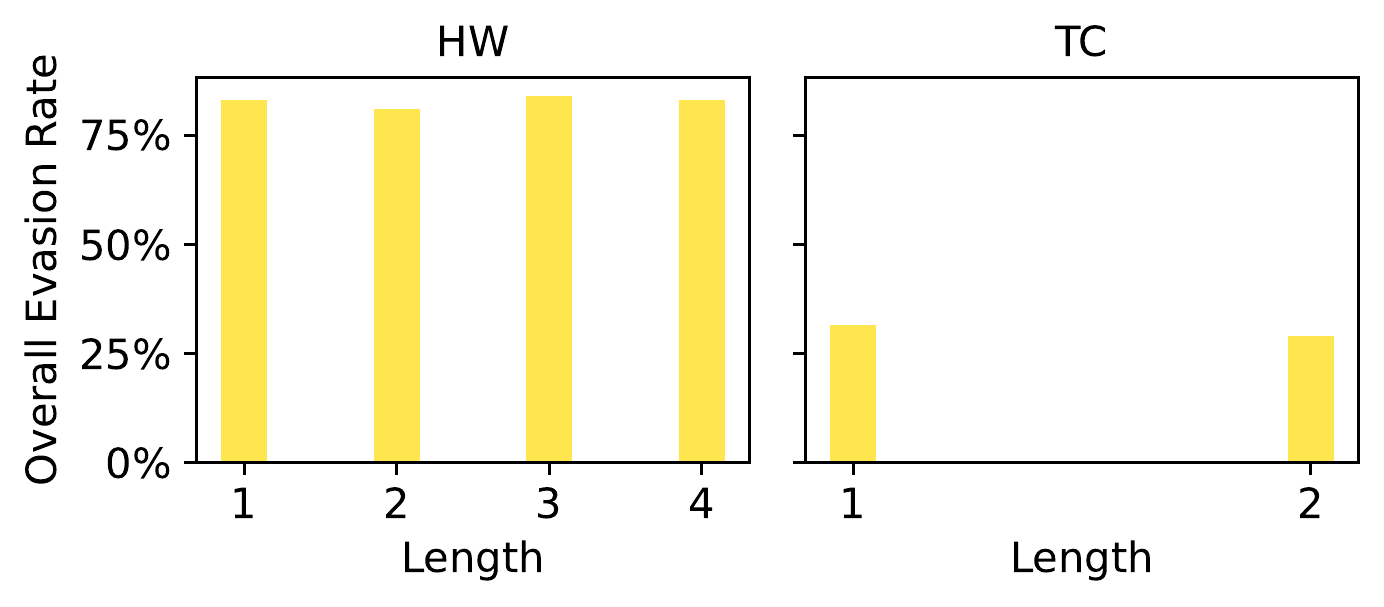}}
\caption{\footnotesize Evaluation of the length of action sequence.}
	
	\label{fig:action_length_evaluation}
\end{figure}

\textbf{Length of Action Sequence.} Intuitively, increasing the length of the action sequence should bring better security gain to action-based FLV.  Under each length of action sequence that the target API supports, we randomly sample from the supported actions to form the action sequence and then utilize \texttt{LiveBugger} to evaluate the target API. We present the evaluation results in Figure \ref{fig:action_length_evaluation}.

From Figure \ref{fig:action_length_evaluation}, we can see that the security of the action-based FLV API is insensitive to the action sequence length. 
For example, as shown by the result on HW, with the length of action sequence increased from 1 to 4, the overall evasion rate is kept at around 80\%. 

 \section{Other Experimental Results}
  \label{appendix_other}

 \begin{figure}[ht]
	\centering
	
	\subfigure[Shaking Head]{
		\begin{minipage}[t]{0.3
		\columnwidth}
			\centering
			\includegraphics[width=\columnwidth]{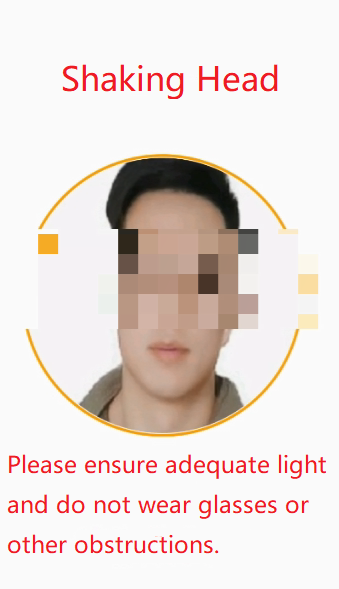}\\
		     \label{screenshot_shakehead}
		\end{minipage}%
	}%
	\subfigure[Pass]{
		\begin{minipage}[t]{0.3\columnwidth}
			\centering
			\includegraphics[width=\columnwidth]{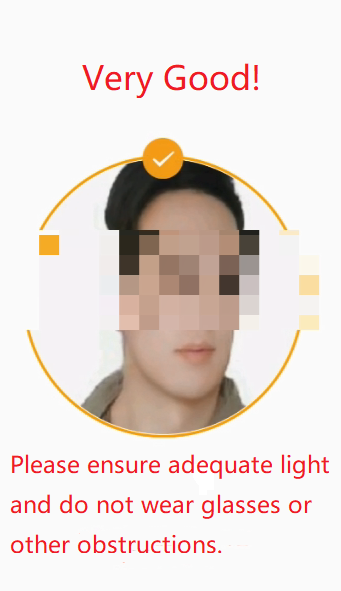}\\
		\label{screenshot_pass}	
		\end{minipage}%
	}%
	\centering
	\caption{\footnotesize Screenshots for evaluating HN Airlines.}

	\label{fig:hainan}
\end{figure}

\textbf{Proof-of-concept Attack.} To better understand the process of a proof-of-concept attack, we take HN Airlines as an example and show the attack screenshot in Figure \ref{fig:hainan}.

\textbf{Evaluation of the Presentation Attacks.} For analyzing the vulnerability of FLV against the deepfake-powered attacks, we also evaluate the effectiveness of the presentation attacks.  We randomly select the replayed images or videos to evaluate the corresponding FLV services and present the results in Table \ref{presentation_attack}. It can be seen that the presentation attacks can hardly pose any threat to the current FLV APIs. 

\begin{table}[h]{\footnotesize
\setlength{\tabcolsep}{2pt}
\centering


\begin{tabular}{ccccc}
FLV                                &         & \begin{tabular}[c]{@{}c@{}}Liveness \\Evasion \end{tabular} & \begin{tabular}[c]{@{}c@{}}Anti-deepfake\\  Evasion \end{tabular} & \begin{tabular}[c]{@{}c@{}}Overall\\  Evasion \end{tabular} \\ \hline
\multirow{2}{*}{Image-based FLV}   & BD   & 2.5\%                                                           & 100\%                                                                & 2.5\%                                                          \\
                                   & TC & 0                                                               & 1.5\%                                                                & 0                                                              \\
\multirow{2}{*}{Silence-based FLV} & BD   & 4.9\%                                                           & 100\%                                                                & 4.9\%                                                          \\
                                   & TC & 2.5\%                                                           & 19\%                                                                 & 0                                                              \\ 
\end{tabular}
\caption{\footnotesize Evaluation of the presentation attacks.}
\label{presentation_attack}}
\end{table}




\end{document}